Article

# Elastocaloric determination of the phase diagram of Sr$_2$RuO$_4$

You-Sheng Li[1], Markus Garst[2,3], Jörg Schmalian[3,4], Sayak Ghosh[5], Naoki Kikugawa[6], Dmitry A. Sokolov[1], Clifford W. Hicks[1,7], Fabian Jerzembeck[1], Matthias S. Ikeda[8,9,10], Zhenhai Hu[1], B. J. Ramshaw[5], Andreas W. Rost[11,12], Michael Nicklas[1]✉ & Andrew P. Mackenzie[1,11]✉

One of the main developments in unconventional superconductivity in the past two decades has been the discovery that most unconventional superconductors form phase diagrams that also contain other strongly correlated states. Many systems of interest are therefore close to more than one instability, and tuning between the resultant ordered phases is the subject of intense research[1]. In recent years, uniaxial pressure applied using piezoelectric-based devices has been shown to be a particularly versatile new method of tuning[2,3], leading to experiments that have advanced our understanding of the fascinating unconventional superconductor Sr$_2$RuO$_4$ (refs. [4–9]). Here we map out its phase diagram using high-precision measurements of the elastocaloric effect in what we believe to be the first such study including both the normal and the superconducting states. We observe a strong entropy quench on entering the superconducting state, in excellent agreement with a model calculation for pairing at the Van Hove point, and obtain a quantitative estimate of the entropy change associated with entry to a magnetic state that is observed in proximity to the superconductivity. The phase diagram is intriguing both for its similarity to those seen in other families of unconventional superconductors and for extra features unique, so far, to Sr$_2$RuO$_4$.

To establish the phase diagram of an unconventional superconductor, it is necessary to have both an effective means of tuning it and methods to investigate the resultant changes to its physical properties. In most of the systems studied so far, tuning methods such as chemical composition, magnetic field, electric field, hydrostatic and epitaxial pressure have been used. Each has its advantages and drawbacks, which ultimately determine the methods used to study the resultant phases and their interplay. An ideal method of study is one that has sensitivity to several phases simultaneously, and in particular to their boundaries. In magnetically tuned systems, the magnetocaloric effect has proved to be of particular utility. Under adiabatic conditions, the rate of change of the sample temperature with applied magnetic field $H$ provides direct information on the heat capacity $C_H$ and entropy $S$, through the well-known relationship

$$\left.\frac{\Delta T}{\Delta H}\right|_S \cong -\frac{T}{C_H}\left.\frac{\partial S}{\partial H}\right|_T, \quad (1)$$

which has been used to good effect to establish the $H$–$T$ phase diagrams of, for example, URu$_2$Si$_2$ (ref. [10]) and Sr$_3$Ru$_2$O$_7$ (ref. [11]). As a tuning parameter, magnetic field brings advantages in terms of directionality and the ability to change symmetry, but also has the clear disadvantage that sufficiently high fields usually destroy, rather than promote, superconductivity. Uniaxial pressure brings the same advantage in terms of 'selective symmetry breaking' and does not automatically compete with superconductivity. In systems with a strong elastic response, the elastocaloric effect is a direct analogue of the magnetocaloric effect:

$$\left.\frac{\Delta T}{\Delta \varepsilon}\right|_S \cong -\frac{T}{C_\varepsilon}\left.\frac{\partial S}{\partial \varepsilon}\right|_T \quad (2)$$

in which $C_\varepsilon$ is the specific heat at constant strain $\varepsilon$ and $\left.\frac{\partial S}{\partial \varepsilon}\right|_T$ is the strain derivative of the entropy at constant temperature. In the special case of an isotropic volume strain $\Delta\varepsilon = \Delta V/V$, then $-\frac{1}{T}\frac{\Delta T}{\Delta\varepsilon}$ measures the famous Grüneisen parameter $\Gamma$ originally introduced in 1908 (ref. [12]) and extensively studied in, for example, heavy fermion materials[13], but a generalized version of the Grüneisen parameter can also be defined for any combination of strain tensor components. If the relevant strain tunes the material through a quantum phase transition, the appropriate generalized Grüneisen parameter is an excellent tool with which to classify that transition[14,15].

[1]Max Planck Institute for Chemical Physics of Solids, Dresden, Germany. [2]Institut für Theoretische Festkörperphysik, Karlsruher Institut für Technologie, Karlsruhe, Germany. [3]Institut für QuantenMaterialien und Technologien, Karlsruher Institut für Technologie, Karlsruhe, Germany. [4]Institut für Theorie der Kondensierten Materie, Karlsruher Institut für Technologie, Karlsruhe, Germany. [5]Laboratory of Atomic and Solid State Physics, Cornell University, Ithaca, NY, USA. [6]National Institute for Materials Science, Tsukuba, Japan. [7]School of Physics and Astronomy, University of Birmingham, Birmingham, UK. [8]Geballe Laboratory for Advanced Materials, Stanford University, Stanford, CA, USA. [9]Department of Applied Physics, Stanford University, Stanford, CA, USA. [10]Stanford Institute for Materials and Energy Sciences, SLAC National Accelerator Laboratory, Menlo Park, CA, USA. [11]Scottish Universities Physics Alliance, School of Physics and Astronomy, University of St Andrews, St Andrews, UK. [12]Max Planck Institute for Solid State Research, Stuttgart, Germany. ✉e-mail: Michael.Nicklas@cpfs.mpg.de; Andy.Mackenzie@cpfs.mpg.de

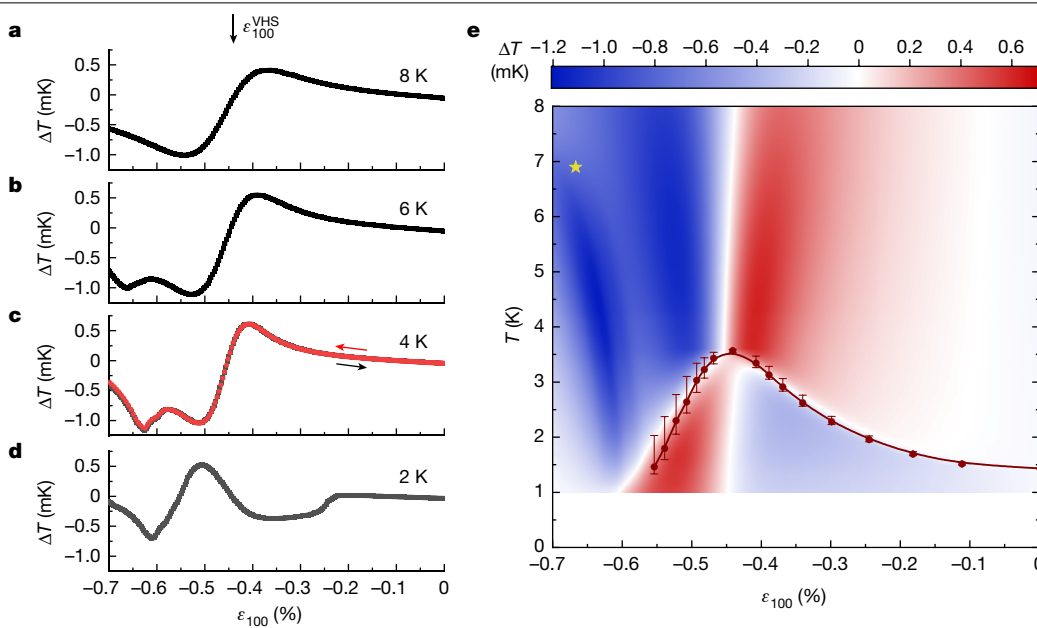

**Fig. 1 | Response of the elastocaloric effect as a function of strain. a–d,** The magnitude of the measured a.c. temperature against strain $\varepsilon_{100}$ at different average sample temperatures, measured at 1,513 Hz and an excitation amplitude $\varepsilon_{100}^{exc}$ between $2.9 \times 10^{-6}$ and $3.5 \times 10^{-6}$. The strain at which the Van Hove singularity is traversed $\varepsilon_{100}^{VHS} = -0.44\%$ is indicated in panel **a**. The sign change of $\Delta T$ at $\varepsilon_{100}^{VHS}$ corresponds to a maximum in the entropy. In panel **c**, data are shown for downsweeps and upsweeps at a rate of approximately 1% per hour. **e,** Colour map of the elastocaloric effect. Notice the pronounced entropy quench at $T_c$, at which the entropy changes from being maximal at $\varepsilon_{100}^{VHS}$ for $T > T_c$ to forming a minimum at $T < T_c$. The solid red circles are the superconducting transition temperatures determined from measurements of the heat capacity[30]. The yellow star indicates the magnetic phase transition temperature obtained from µSR data[23], which agrees with the phase boundary identified by the dark blue contrast seen for $\varepsilon_{100}$ between −0.6% and −0.7% in the elastocaloric effect. See Extended Data Figs. 1, 2 and 3 for further data.

Although used widely in association with materials with large elastic responses, to the extent that it has been proposed for cooling technologies[16,17], direct measurement of the elastocaloric effect has been much less widely used in the field of unconventional superconductivity or correlated electron physics, partly because the expected signal size is much smaller. Here we build on recent work using a.c. methods to perform high-resolution measurements of $\Delta T/\Delta\varepsilon$ in Fe-based superconductors[18–20] to study the elastocaloric effect in $Sr_2RuO_4$. As described in detail in Methods, we superimpose a small oscillatory component on the background steady strain and lock into the oscillatory component of the thermal response, which directly measures $\Delta T/\Delta\varepsilon$. We achieve the extremely high temperature measurement precision of approximately 2 µK (√Hz)$^{-1}$ and use it to map out the phase diagram between 1 K and 8 K, for applied compressive strains along the [100] crystal axis of up to $\varepsilon_{100} = -0.7\%$, performing checks to ensure that we are close to the adiabatic limit for which Equation (2) applies. Our data allow us to determine $\Gamma_{100}$, the Grüneisen parameter for uniaxial stress applied along [100].

Sample raw data for isothermal strain sweeps at 8 K, 6 K, 4 K and 2 K are shown in Fig. 1a–d. Much can be learned from a qualitative inspection of the results. At 8 K, the data are seen to show the profile expected for a system in which a peak in entropy is studied under quasi-adiabatic conditions: the derivative changes sign at $\varepsilon_{100} \cong -0.44\%$, in line with previous estimates for the strain at which a Van Hove singularity is traversed at a so-called Lifshitz transition[5,8,21,22].

When the temperature has been lowered to 6 K, the signal at the Van Hove strain $\varepsilon_{100}^{VHS}$ remains similar, but a pronounced extra dip is seen in the signal at around $\varepsilon_{100} \approx -0.6\%$. By 4 K, this dip has moved to slightly lower absolute strain and becomes stronger. The signal at 2 K looks similar to those at 4 K and 6 K at high strain but is very different in the region between −0.2% and −0.6% strain. Instead of a maximum of entropy at the Van Hove strain, there is now a minimum, along with a sharp step in the elastocaloric signal at $\varepsilon_{100} \cong -0.23\%$. Remarkably, this large change in the entropic properties is the result of the onset of superconductivity, as demonstrated by constructing the empirical phase diagram shown in Fig. 1e from interpolating the results of strain sweeps from 71 different temperatures, as described in Methods.

The high resolution of our experiments allows the straightforward identification of several key features from inspection of the raw data in Fig. 1e. First, above $T_c$, the strain at which the elastocaloric signal changes sign is nearly independent of temperature. This is the intuitive expectation for the elastocaloric signal of traversing a Van Hove singularity, which is expected to be independent of temperature in this temperature range because it is set by an underlying feature in the band structure and therefore determined by much higher energy scales. Within experimental uncertainty, it coincides with the maximum value of $T_c$ and the strain at which the Van Hove singularity is observed to be crossed in photoemission experiments[8]. Second, the dispersion with strain of the dip seen for $\varepsilon_{100}$ values of less than −0.6% (Fig. 1b–d) is reminiscent of that of a phase boundary. Third, the entropic signal of entering the superconducting state is extremely pronounced. The maximum in entropy as a function of strain at the Van Hove singularity is quenched, turning into a minimum below $T_c$. Away from the Van Hove point, the elastocaloric effect changes sign and almost reverses its magnitude near $T_c$. No signature of a second transition within the superconducting state[23] is resolved.

To frame a more in-depth analysis of our data, we turn to the behaviour of the relevant Grüneisen parameter $\Gamma_{100} \equiv -\frac{1}{T}\frac{\Delta T}{\Delta\varepsilon_{100}}$, converting the raw data to absolute units using the procedure described in Methods. In systems governed by a single energy scale, such as Fermi liquids, $\Gamma$ is independent of temperature. As a result, Grüneisen scaling is expected with curves at all temperatures collapsing onto each other and deviations from this scaling indicating proximity to critical points or phase transitions[14,15]. We show this scaling in Fig. 2a for temperatures greater than the maximum superconducting transition temperature of 3.5 K. It is seen to be excellent for $-0.3\% < \varepsilon_{100} < 0$. Between −0.3% and −0.55%, the departure from scaling is of the kind qualitatively

Article

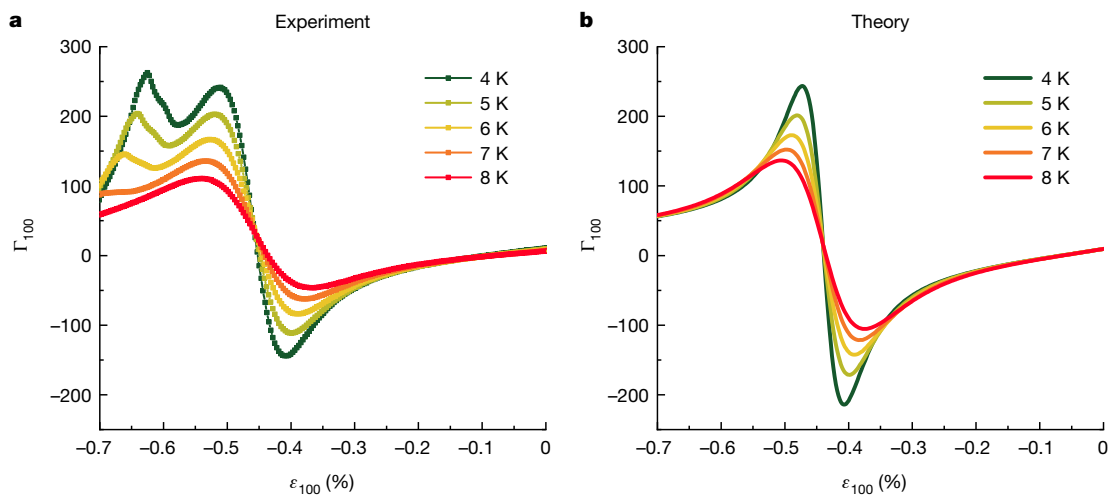

**Fig. 2 | Grüneisen scaling. a**, Experimental data converted to absolute units using a temperature-independent and strain-independent scale factor determined at 6 K using the procedure described in Methods. **b**, Theoretical calculations of $\Gamma_{100}$ in the temperature range $T \geq 4$ K using a simple single-band, two-dimensional model. The sign change at $\varepsilon_{100} \approx -0.44\%$ can clearly be attributed to crossing the Van Hove point. A Grüneisen scaling collapse is observed for small strains $\varepsilon_{100} > -0.2\%$ in both panels. The extra peak at large strains for $\varepsilon_{100}$ around $-0.65\%$ in panel **a** is attributed to magnetism that is not captured by the theory of panel **b**. A more realistic model including the full three-dimensional dispersion of $Sr_2RuO_4$ (Supplementary Information) gives essentially the same theoretical results.

expected for proximity to a quantum phase transition, in this case, the Lifshitz transition at $\varepsilon_{100} = -0.44\%$. For strains between $-0.6\%$ and $-0.7\%$, the Grüneisen scaling is also poorly obeyed, supporting the hypothesis that the feature in this region (now a peak rather than a dip because of the sign convention of the Grüneisen parameter) marks a phase transition.

As a complement to the experimental data, we have calculated the expected behaviour of the Grüneisen parameter as a function of [100] strain using a two-dimensional tight-binding model for the relevant $\gamma$ band derived from a combination of de Haas–van Alphen and angle-resolved photoemission experiments on unstrained $Sr_2RuO_4$ (ref. [24]). Full details are given in Methods and Supplementary Information.

The results are shown in Fig. 2b for the same range of temperatures as those in Fig. 2a. The qualitative agreement at strains $|\varepsilon_{100}| < |\varepsilon_{100}^{VHS}|$ is distinctive, especially given the simplicity of the model. The shape of the curves, the strain range over which the Grüneisen scaling is obeyed and even the zero crossing near zero strain (a consequence of the initial splitting of the zero-field Van Hove singularity owing to the Poisson effect) are all seen in both experiment and theory. By contrast, below 8 K, the behaviour for strains beyond $\varepsilon_{100}^{VHS}$ is considerably different, emphasizing that the experimental data are picking up a phase transition not predicted by the tight-binding model. In isolation, the elastocaloric data give no microscopic information on the nature of the high-strain phase, but a point established in a recent muon spin relaxation measurement (marked by the yellow star in Fig. 1e) shows that it is magnetic, and probably a finite-$Q$ state[23]. Establishing the boundary of this new phase is one of our key findings.

Next, we turn our attention to lower temperatures. In Fig. 3a, we show elastocaloric data at a range of temperatures between 3.7 K and 1 K. At 3.7 K, the sample is non-superconducting across the entire strain range, whereas at 1 K, it is superconducting for $\varepsilon_{100}$ between $-0.55\%$ and 0. The behaviour at intermediate temperatures is prominent. After following the expected normal state behaviour at low strains, the signal abruptly reverses in sign owing to the entropy quench discussed above. On the high-strain side, the departure from the superconducting state becomes harder to distinguish and the signal increases rapidly in the region where superconducting and magnetic order approach each other.

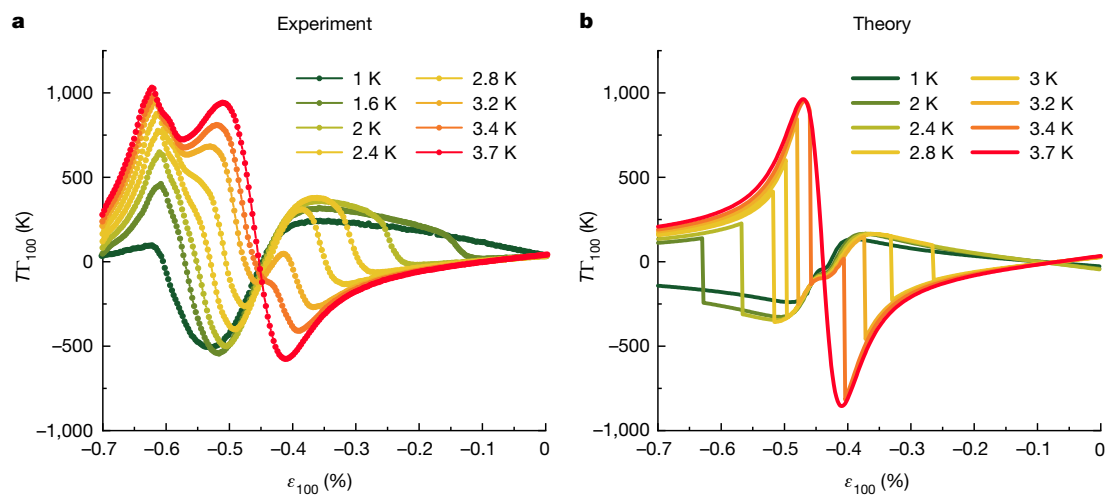

**Fig. 3 | Elastocaloric effect. a**, Experimental data converted to absolute units using the procedure described in the caption to Fig. 2. **b**, Theoretical calculations with a full gap at the Van Hove point of $T\Gamma_{100}$ in the temperature range $1\,\text{K} \leq T \leq 3.7\,\text{K}$ using a simple single-band, two-dimensional model. The discontinuities in panel **b** identify the phase transition between the metallic and superconducting phases; these singularities are broadened in the experimental data in panel **a**. The extra peak at large strains for $\varepsilon_{100}$ around $-0.65\%$ is attributed to magnetism.

Because there is, to our knowledge, no precedent in the literature of measurement of the elastocaloric signal on entry to the superconducting state, we constructed an illustrative model to frame the discussion of Fig. 3a. In Fig. 3b, we show the elastocaloric response obtained in a simple calculation: the density of states of the empirically constrained tight-binding model is combined with a strain-independent and $k$-independent pairing potential $V$ to calculate the transition temperature of a hypothetical weak-coupled Bardeen–Cooper–Schrieffer superconductor that is fully gapped at the Van Hove points (see Supplementary Information). We do not claim that this model gives a full description of the superconductivity of $Sr_2RuO_4$ and certainly do not expect it to accurately predict $T_c(\varepsilon_{100})$ across the entire strain range, but it usefully highlights some of the key features of the experimental data. It demonstrates that the pronounced signal sign reversal that is so visually prominent in Fig. 3a for $-0.35\% < \varepsilon_{100} < -0.1\%$ on entry to the superconducting state can be understood within a very simple model of superconductivity. Consistent with the trend seen in the data, the large entropy at the Van Hove singularity arising from the enhanced density of states is strongly quenched on entering the superconducting state. By contrast, a second model calculation shows that our data cannot be reproduced by superconducting states with nodes at the Van Hove points (see Supplementary Information). Our data are therefore consistent only with superconducting order parameters that give a substantial gap in the vicinity of the Van Hove singularity. Nodal lines or points away from the Van Hove point are, of course, still possible.

Arguably as interesting as what the model successfully describes is what it does not. The theory–experiment comparison in Fig. 3 again highlights the notable qualitative difference in the experimental data on the low-strain and high-strain sides of $\varepsilon_{100}^{VHS}$. The models used to construct Figs. 2b and 3b do not include provision for a magnetic phase at high strain and predict a high degree of symmetry of both the normal state and the superconducting state signals around $\varepsilon_{100}^{VHS}$. The pronounced asymmetry in the data shows that the magnetic state exists and suggests that it affects the superconductivity.

It is possible to go further than qualitative statements and to extract the strain dependence of the entropy, using the analysis procedure described in Methods. Sample results at 4.5 K, 5.5 K, 6.5 K and 7.5 K are shown in Fig. 4. After peaking at the Van Hove strain, the entropy decreases as the strain is increased, with a more sudden decrease for $-0.61\% < \varepsilon_{100} < -0.68\%$, whose magnitude increases with decreasing temperature. At a first-order phase transition, the entropy shows a discontinuity, whereas we observe instead a rapid decrease of finite width. However, our experiment involves a small strain inhomogeneity, whose effects are clearly seen in Fig. 3 in the broadening of the signal as the superconducting state is entered. The strain width of the entropy decrease in Fig. 4 is similar, so the data probably indicate that the intrinsic decrease is discontinuous. The raw elastocaloric data highlight the qualitative difference between the signature of a peaking entropy (seen at $\varepsilon_{100}^{VHS}$) and the signature seen on entering the magnetic phase, which is a peak not in the entropy but in $-(\partial S/\partial \varepsilon)_T$. Overall, although we cannot be absolutely certain, we believe that our data support a first-order transition into the magnetic phase. Taking the peak in $\Gamma_{100}$ as the transition point identifies it with the dark blue ridge in Fig. 1e. We can also quantify the change of entropy. The decrease in $S/T$ at 4.5 K is approximately 3 mJ mol$^{-1}$ K$^{-2}$, 8% of the electronic entropy of the unstrained material and more than 10% of the extrapolated background value at $\varepsilon_{100} = -0.63\%$. The absolute value is similar to that seen on entry to the low-temperature phase in $Sr_3Ru_2O_7$ (ref. [11]) but the sign is opposite. In $Sr_2RuO_4$, the entropy is lower in the magnetic phase than in the adjacent metal, in line with conventional expectation for a Fermi surface gapping transition.

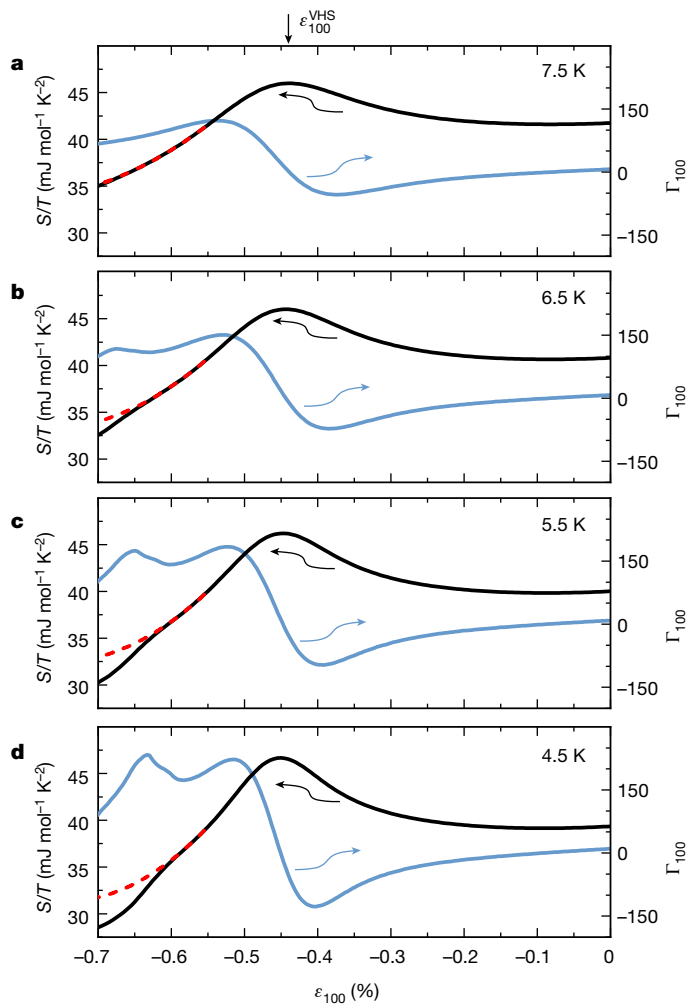

**Fig. 4 | Strain dependence of the entropy. a–d**, Integrating the $\Gamma_{100}$ data using the iterative procedure described in Methods yields the strain-dependent entropy, plotted as $S/T$ (black line) and directly compared with the measured $\Gamma_{100}$ (blue line). The peak in $\Gamma_{100}$ and corresponding rapid decrease of $S/T$ indicates the magnetic transition at 4.5 K, 5.5 K and 6.5 K (**b**–**d**), whereas no decrease is visible at 7.5 K (**a**). The dotted red lines in the entropy traces are extrapolations of the background $S/T$ from −0.55% to a lower cut-off consistent with the varying onset of the magnetic transition (−0.58%, −0.59%, −0.60% and −0.61% for 4.5 K, 5.5 K, 6.5 K and 7.5 K, respectively).

Independent of microscopic detail, the phase diagram determined by our measurements shows a strong and previously unappreciated experimental similarity between the tuned phase diagram of $Sr_2RuO_4$ and those of many cuprate, pnictide, organic and heavy fermion superconductors[25–28], in which superconductivity appears in the vicinity of a magnetic phase that is driven towards zero temperature by an external tuning parameter. This is especially timely given the recent evidence for an even-parity order parameter in $Sr_2RuO_4$ (ref. [9]), in common with those thought to exist in most of the above-mentioned materials. Our experiments also highlight the utility of the a.c. elastocaloric effect in the general study of unconventional superconductivity and correlated electron physics. As we have shown, the elastocaloric effect enables rapid and comprehensive phase diagram mapping and provides high-resolution datasets against which the quantitative predictions of theory can be tested. In the uniaxial pressure apparatus that we have developed[3,29], its combination with other experiments should be fairly straightforward, offering simultaneous access to spectroscopic and thermodynamic information from the same sample.

# Article

## Online content

Any methods, additional references, Nature Research reporting summaries, source data, extended data, supplementary information, acknowledgements, peer review information; details of author contributions and competing interests; and statements of data and code availability are available at https://doi.org/10.1038/s41586-022-04820-z.

**Publisher's note** Springer Nature remains neutral with regard to jurisdictional claims in published maps and institutional affiliations.

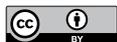

## Methods

### Sample preparation and experimental setup

High-quality single crystals of $Sr_2RuO_4$ were grown by a floating-zone method[31]. Special care was taken to select the sample from a region with the highest $T_c$ and the absence of a signal from '3-K phase' inclusions, indicating the highest quality $Sr_2RuO_4$. The sample was aligned along the [100] crystallographic direction by Laue X-ray diffraction and needles were wire sawed and polished using diamond-impregnated sheets with different grain sizes down to 1 μm to obtain parallel surfaces and to reduce the surface roughness. A home-made Au/AuFe (0.07%) thermocouple (25 μm wire diameter) served as a thermometer to measure the a.c. temperature changes. It was independently calibrated using procedures outlined in ref. [30] and attached to the centre of the sample using silver epoxy (Dupont 6838), soldered to twisted copper wires that were thermally anchored on the thermometry stage. The assembly was subsequently glued in the jaws of a uniaxial pressure cell using Stycast 2850FT epoxy with Catalyst 23LV. Special care was taken to minimize the tilt of the sample and to ensure a force transmission along the long axis of needle. The sample temperature was measured using calibrated resistive low-temperature sensors. The present experimental setup with the same sample was used in a previous heat capacity study[30,32]. Schematic diagrams of the experimental setup are shown in Extended Data Fig. 4a,b.

### Measurement of the elastocaloric effect

The elastocaloric effect was measured by an a.c. modulation method[18]. The uniaxial pressure apparatus was mounted to the cold plate of a dilution refrigerator (Oxford Instruments). To achieve the large strains needed to tune $Sr_2RuO_4$ in the desired range, large d.c. voltages had to be applied on both inner and outer piezoactuators of the uniaxial pressure apparatus. A home-made high-voltage amplifier was used to drive the outer piezoactuators. The a.c.-modulated strain was achieved by superimposing an a.c. voltage on top of a d.c. voltage on the inner piezoactuator. To amplify the coupled a.c. and d.c. voltages, a commercial high-voltage amplifier was used (TEGAM 2350, bandwidth d.c. to 2 MHz). The extremely low noise level of 20 pV (√Hz)$^{-1}$ on the thermocouple readout, corresponding to 5.1 μK (√Hz)$^{-1}$, 2.1 μK (√Hz)$^{-1}$ and 1.7 μK (√Hz)$^{-1}$ at 1 K, 4 K and 8 K, respectively, was obtained by the use of a high-frequency low-temperature transformer (CMR-Direct), operating at a gain of 300, mounted on the 1-K pot of the dilution refrigerator. Its output was read by an EG&G 7265 lock-in amplifier. We show the configuration of the electronic setup for the ultra-low-noise measurement of the temperature oscillations in Extended Data Fig. 4c.

### Determination of the applied uniaxial strain in the sample

Strain is the change of the length of a sample $\Delta l = l - l_0$ divided by its length $l_0$. The strain apparatus used in this study has a capacitor to measure the displacement $\Delta d$ obtained by applying a voltage to the piezoelectric actuators (PEAs). However, the measured $\Delta d$ is not the change in the sample length. $\Delta l$ can be obtained by the change of the capacitor displacement $\Delta d$ times a transfer efficiency $e$, which is defined by the properties of Stycast layers between the sample and the jaws of the strain apparatus[3]. Therefore, we find for the strain in the sample:

$$\varepsilon = \frac{\Delta l}{l} = \frac{e \times \Delta d}{l_0}. \quad (3)$$

In the case of $Sr_2RuO_4$ in the current setup, a transfer efficiency $e = 0.78$ could be estimated on the basis of the known position of the maximum in $T_c$ for an applied stress along [100] at 0.7 GPa (ref. [29]) and the Young's modulus $E_Y = 160$ GPa at 4 K (ref. [33]).

To obtain the large strains needed to investigate the phase diagram of $Sr_2RuO_4$, the inner and outer PEAs of the strain apparatus are used. To measure the elastocaloric effect, a further small a.c. voltage is imposed on the d.c. voltage applied on the inner PEA. The oscillation amplitude $d_{exc}$ can be measured using the capacitor mounted in parallel to the sample and the strain amplitude is then obtained following Equation (3):

$$\Delta\varepsilon = \frac{e \times d_{exc}}{l_0}. \quad (4)$$

In our case, the displacement amplitude $d_{exc}$ is between 5 nm and 10 nm, in comparison with a sample length of approximately 2 mm. Strain is a tensor quantity, so a formal definition of $\varepsilon_{100}$ as used in the main text is $\varepsilon_{100} = \vec{e}_{100} \cdot \hat{\varepsilon} \cdot \vec{e}_{100}$, in which $\vec{e}_{100} = (1,0,0)$.

### Adiabaticity of the measurement

Curves of $\Delta T$ against frequency at 0.5% compression are shown in Extended Data Fig. 5 on a double-logarithmic representation. One can easily identify the lower cut-off frequency, between 100 Hz and 300 Hz. In the high-frequency range, this is not possible because the data start to scatter strongly above a few kilohertz before the upper cut-off frequency is reached. The enhanced noise is related to vibrations of thermocouple wires. Between 1 K and 8 K, we do not observe a notable change in the upper frequency boundary. This implies that the upper cut-off frequency is at least larger than 10 kHz. Here we chose a measuring frequency $f = 1{,}513$ Hz, which corresponds to $\Delta T$ on the plateau of the frequency response. The phase response is around zero for all temperatures between 1 K and 8 K at $f = 1{,}513$ Hz.

### Estimation of the elastocaloric signal size

In principle, the absolute value of the elastocaloric effect can be obtained directly. However, owing to the smallness of the signal and uncertainties arising from sample configuration and material properties, it is more reliable to calibrate the elastocaloric effect $\Delta T_{ad}/\Delta\varepsilon$, as described in the following.

The elastocaloric effect can be described as an adiabatic temperature change $\Delta T_{ad}$ as a function of strain $\varepsilon$:

$$\frac{\Delta T_{ad}}{\Delta\varepsilon} \cong -\frac{T}{C_{\varepsilon,\sigma_y,\sigma_z}} \left(\frac{\partial S}{\partial \varepsilon}\right)_{T,\sigma_y,\sigma_z} \quad (5)$$

Here $C_{\varepsilon,\sigma_y,\sigma_z}$ is the heat capacity at constant strain and $S$ is the entropy. The relevant elastocaloric Grüneisen parameter $\Gamma$ in our experiment is related to entropy through

$$\Gamma = \frac{(\partial S/\partial\varepsilon)_{T,\sigma_y,\sigma_z}}{C_{\varepsilon,\sigma_y,\sigma_z}} = \frac{(\partial S/\partial\varepsilon)_{T,\sigma_y,\sigma_z}}{T(\partial S/\partial T)_{\varepsilon,\sigma_y,\sigma_z}} = -\frac{1}{T}\left(\frac{\partial T}{\partial\varepsilon}\right)_{S,\sigma_y,\sigma_z} \quad (6)$$

Please note that, throughout this section, $\varepsilon$ refers to $\varepsilon_{xx}$. At very low strains on the order of $-0.1\%$ at temperatures above the superconducting transition, one can treat the system as being a Fermi liquid whose parameters are a function of strain. In this case, the specific heat to the second order in strain is given by

$$C(\varepsilon, T) = \gamma\left(1 + \varepsilon\gamma_1/\gamma + \varepsilon^2\gamma_2/\gamma\right)T + \beta T^3 \quad (7)$$

Here we further assumed that the phonon heat capacity in our case has a negligible strain dependence. This is justified by both the small strain limit considered and the fact that the phonon contribution is much smaller than the electronic heat capacity at the relevant temperatures in the first place. It directly follows that entropy $S$ is given by

$$S(\varepsilon, T) = \int_0^T \frac{C(\varepsilon, T)}{T} dT = (\gamma + \gamma_1\varepsilon + \gamma_2\varepsilon^2)T + \frac{1}{3}\beta T^3 \quad (8)$$

# Article

In this limit, the elastocaloric Grüneisen parameter Γ can be expressed as

$$\Gamma = \frac{(\gamma_1 + 2\gamma_2 \varepsilon) T}{(\gamma + \gamma_1 \varepsilon + \gamma_2 \varepsilon^2) T + \beta T^3} \quad (9)$$

and

$$(\partial S/\partial \varepsilon)_{T,\sigma_y,\sigma_z} = (\gamma_1 + 2\gamma_2 \varepsilon) T \quad (10)$$

Furthermore, one can consider the second derivative of entropy with respect to strain

$$\left(\frac{\partial^2}{\partial \varepsilon^2} S\right)_{T,\sigma_y,\sigma_z} = \left(\frac{\partial}{\partial \varepsilon}\left(\frac{\partial S}{\partial \varepsilon}\right)_{T,\sigma_y,\sigma_z}\right)_{T,\sigma_y,\sigma_z} = -\left(\frac{\partial}{\partial \varepsilon}\left(\frac{\partial \sigma_x}{\partial T}\right)_{\varepsilon,\sigma_y,\sigma_z}\right)_{T,\sigma_y,\sigma_z} \quad (11)$$

in which we made use of the appropriate Maxwell relationship in the last step. Given that in the range considered here thermodynamic variables are well behaved, it follows that

$$\left(\frac{\partial^2}{\partial \varepsilon^2} S\right)_{T,\sigma_y,\sigma_z} = -\left(\frac{\partial}{\partial T}\left(\frac{\partial \sigma_x}{\partial \varepsilon}\right)_{T,\sigma_y,\sigma_z}\right)_{\varepsilon,\sigma_y,\sigma_z} \quad (12)$$

in which stress $\varepsilon$ and strain $\sigma$ are related by means of the compliance matrix $\underline{s}$ through $\boldsymbol{\varepsilon} = \underline{s}\boldsymbol{\sigma}$. Hence

$$\left(\frac{\partial^2}{\partial \varepsilon^2} S\right)_{T,\sigma_y,\sigma_z} = -\left(\frac{\partial}{\partial T} s_{11}^{-1}\right)_{\varepsilon,\sigma_y,\sigma_z} \quad (13)$$

with $s_{11}$ being the 11 entry of $\underline{s}$ and the inverse of the Young's modulus.

Combining Equations (10) and (13) therefore yields

$$2\gamma_2 T = -\left(\frac{\partial}{\partial T} s_{11}^{-1}\right)_{\varepsilon,\sigma_y,\sigma_z} \quad (14)$$

$s_{11}$ determined from resonant ultrasound experiments using methods described in ref. [33] is shown in Extended Data Fig. 6a, together with a fit of the form

$$s_{11} = s_{11,0} + s_{11,2} T^2 \quad (15)$$

giving $s_{11,2} = 1.526 \times 10^{-7}$ GPa$^{-1}$ K$^{-2}$ and $\gamma_2 = \frac{s_{11,2}}{s_{11}^2} \approx \frac{s_{11,2}}{s_{11,0}^2} = 0.0039$ GPa K$^{-2}$.

The elastocaloric Grüneisen parameter is therefore fully determined except for $\gamma_1$, permitting us to calibrate our measured data by means of an overall amplitude factor $\Gamma = a\Gamma^{\text{meas}}$,

$$\Gamma^{\text{meas}} = \frac{1}{a} \frac{(\gamma_1 + 2\gamma_2 \varepsilon) T}{(\gamma + \gamma_1 \varepsilon + \gamma_2 \varepsilon^2) T + \beta T^3} \quad (16)$$

$\gamma$, $\beta$ and $s_{11,2}$ are constrained by independent experiments, with $a$ and $\gamma_1$ being the only independent parameters.

In Extended Data Fig. 6b, we show $\Gamma^{\text{meas}}$ for temperatures between 5.5 K and 6.5 K and small strains for up to −0.1%, for which the above approximations are valid. The surface shown is a fit of the functional form of Equation (16). The fit gives $a = 2.90$ and $\gamma_1 = 6,797$ J m$^{-3}$ K$^{-1}$.

## Numerical calculation of the entropy

Here we describe the numerical scheme for the calculation of entropy shown in Fig. 4.

Our starting point is

$$\Gamma_\varepsilon = \frac{1}{C_{\varepsilon_{xx}}} \left(\frac{\partial S}{\partial \varepsilon_{xx}}\right)_T \quad (17)$$

Simple rearrangement gives

$$(\partial S/\partial \varepsilon_{xx})_T = C_{\varepsilon_{xx}} \Gamma_\varepsilon \quad (18)$$

With the knowledge of the entropy at zero strain, $S(\varepsilon = 0, T)$, one can integrate this partial differential to give

$$S(\varepsilon, T) = \int_0^\varepsilon C_{\varepsilon_{xx}}(\varepsilon', T) \, \Gamma_\varepsilon(\varepsilon', T) \, d\varepsilon' + S(\varepsilon = 0, T) \quad (19)$$

$\Gamma_\varepsilon(\varepsilon, T)$ is known from the experiments, whereas $C_{\varepsilon_{xx}}(\varepsilon, T)$ is, at this stage, unknown. However, it is related to $S(\varepsilon, T)$ through

$$C_{\varepsilon_{xx}}(\varepsilon', T) = T \left(\frac{\partial S}{\partial T}\right)_{\varepsilon_{xx}} \quad (20)$$

We therefore use an iterative scheme with the following steps. First, set $C_{\varepsilon_{xx}}^{(0)}(\varepsilon_{xx}, T) = C(0, T)$. This is the zero-strain specific heat, known with very high accuracy for Sr$_2$RuO$_4$. Second, calculate $S^{(0)}(\varepsilon_{xx}, T)$ using Equation (19) and $C_{\varepsilon_{xx}}^{(0)}(\varepsilon_{xx}, T)$ for all available $T$. Third, calculate $C_{\varepsilon_{xx}}(\varepsilon_{xx}, T)$ by interpolating $S^{(0)}(\varepsilon_{xx}, T)$ as a function of $T$ and evaluating Equation (20). Finally, calculate $S^{(1)}(\varepsilon_{xx}, T)$ using $C_{\varepsilon_{xx}}^{(1)}(\varepsilon_{xx}, T)$ and Equation (19), an iteration of the second step.

After a few iterations, no notable changes in $S^{(n)}(\varepsilon, T)$ are observed. This is not least due to the fact that, although $C_{\varepsilon_{xx}}$ does vary overall, these variations are at most a few tens of per cent, enabling an effective convergence of the above scheme. The data shown in the main text correspond to $S^{(2)}$.

## Theoretical analysis

The theoretical analysis of the elastocaloric effect is on the basis of a quasiparticle description of Sr$_2$RuO$_4$. We use a strain-dependent quasiparticle dispersion $\varepsilon_{\mathbf{k}}(\epsilon_{\alpha\beta})$ and determine the electronic contribution to the entropy of the system from

$$S_{\text{el}} = -\frac{2k_{\text{B}}}{N} \sum_{\mathbf{k}} [f_{\mathbf{k}} \log f_{\mathbf{k}} + (1 - f_{\mathbf{k}}) \log(1 - f_{\mathbf{k}})].$$

$f_{\mathbf{k}}$ is the Fermi distribution function with the above dispersion. The factor 2 refers to the electron spin and the sum goes over the momenta in the first Brillouin zone. The elastocaloric coefficient follows from the temperature and strain derivatives of the entropy. We use the following tight-binding parameterization for the $\gamma$-band of the system as determined from angle-resolved photoemission experiments[34]:

$$\varepsilon_{\mathbf{k}} = -2t_x \cos(k_x a_x) - 2t_y \cos(k_y a_y) - 4t' \cos(k_x a_x)\cos(k_y a_y) - \mu,$$

with $t_x = t_y = t_0 = 0.119$ eV, $t' = 0.392 t_0$ and $\mu = 1.48 t_0$. To describe the strain dependence of $\varepsilon_{\mathbf{k}}(\epsilon_{\alpha\beta})$, we assume a linear dependence of the hopping elements with respect to the interatomic distance. The proportionality factor is chosen to reproduce the strain value at which the Van Hove singularity is reached. In the superconducting state, we use the Bogoliubov quasiparticle dispersion $\varepsilon_{\mathbf{k}} \to \sqrt{\varepsilon_{\mathbf{k}}^2 + \Delta^2}$ with superconducting gap $\Delta$. The strain dependence is dominated by the electronic spectrum near the Van Hove point. In our theory, we consider a pairing state that is fully gapped at the Van Hove momentum. The strain dependence of the superconducting gap amplitude and of the transition temperature follow from the solution of the gap equation at fixed pairing interaction. For details, see the Supplementary Information.

## Data availability

The data that underpin the findings of this study are available at https://doi.org/10.17630/6a4a06c6-38d3-464f-88d1-df8d2dbf1e75.

**Acknowledgements** We are grateful to S. Kivelson for helpful discussions. This work was supported by the Max Planck Society and the Deutsche Forschungsgemeinschaft (DFG, German Research Foundation) – TRR 288-422213477 ELASTO-Q-MAT (projects A10 (C.W.H. and A.P.M.), A11 (M.G.) and B01 (J.S.)). S.G. and B.J.R. acknowledge funding from the U.S. Department of Energy, Office of Basic Energy Sciences under award number DESC0020143. This work made use of the Cornell Center for Materials Research (CCMR) Shared Facilities, which is supported through the NSF MRSEC programme (no. DMR-1719875) (S.G. and B.J.R.). N.K. acknowledges support from KAKENHI Grants-in-Aid for Scientific Research (grant nos. 17H06136, 18K04715 and 21H01033) and Core-to-Core Program (no. JPJSCCA20170002) from the Japan Society for the Promotion of Science (JSPS) and by a JST-Mirai Program (grant no. JPMJMI18A3). A.W.R. acknowledges support from the Engineering and Physical Sciences Research Council (grant numbers EP/P024564/1, EP/S005005/1 and EP/V049410/1). Research in Dresden benefits from the environment created by the DFG Excellence Cluster 'Correlations and Topology in Quantum Materials'.

**Author contributions** Y.-S.L. performed the experiments and analysed the data, with input from M.S.I., M.N., Z.H., A.W.R., C.W.H. and A.P.M.; M.G. and J.S. constructed the theoretical models; S.G. and B.J.R. conducted further resonant ultrasound experiments; A.W.R. and Z.H. designed and tested the entropy calculation and signal calibration protocols; N.K., D.A.S. and F.J. grew and characterized the crystals; C.W.H. developed the uniaxial pressure apparatus. A.P.M. conceived the project and wrote the paper, with input from all co-authors.

**Funding** Open access funding provided by Max Planck Society.

**Competing interests** The authors declare no competing financial interests.

**Additional information**
**Supplementary information** The online version contains supplementary material available at https://doi.org/10.1038/s41586-022-04820-z.
**Correspondence and requests for materials** should be addressed to Michael Nicklas or Andrew P. Mackenzie.
**Peer review information** *Nature* thanks Thierry Klein, Cedric Weber and the other, anonymous, reviewer(s) for their contribution to the peer review of this work.
**Reprints and permissions information** is available at http://www.nature.com/reprints.

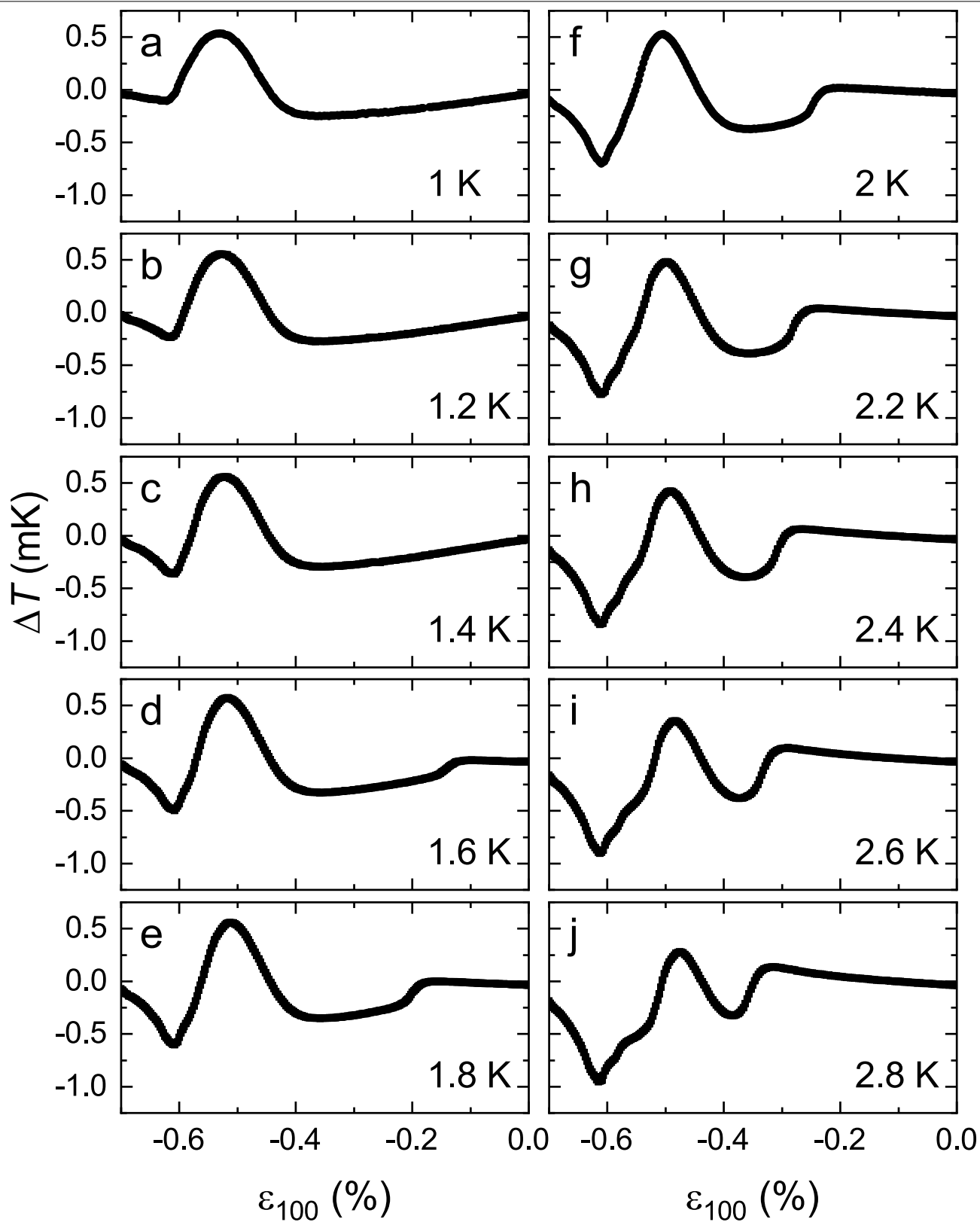

**Extended Data Fig. 1 | Response of the elastocaloric effect as a function of strain (further data: low temperatures).** $\Delta T(\varepsilon_{100})$ recorded at 1,513 Hz with an excitation amplitude $\varepsilon_{100}^{exc}$ between $2.9 \times 10^{-6}$ and $3.5 \times 10^{-6}$ during strain sweeps at ten different temperatures from 1 K to 2.8 K.

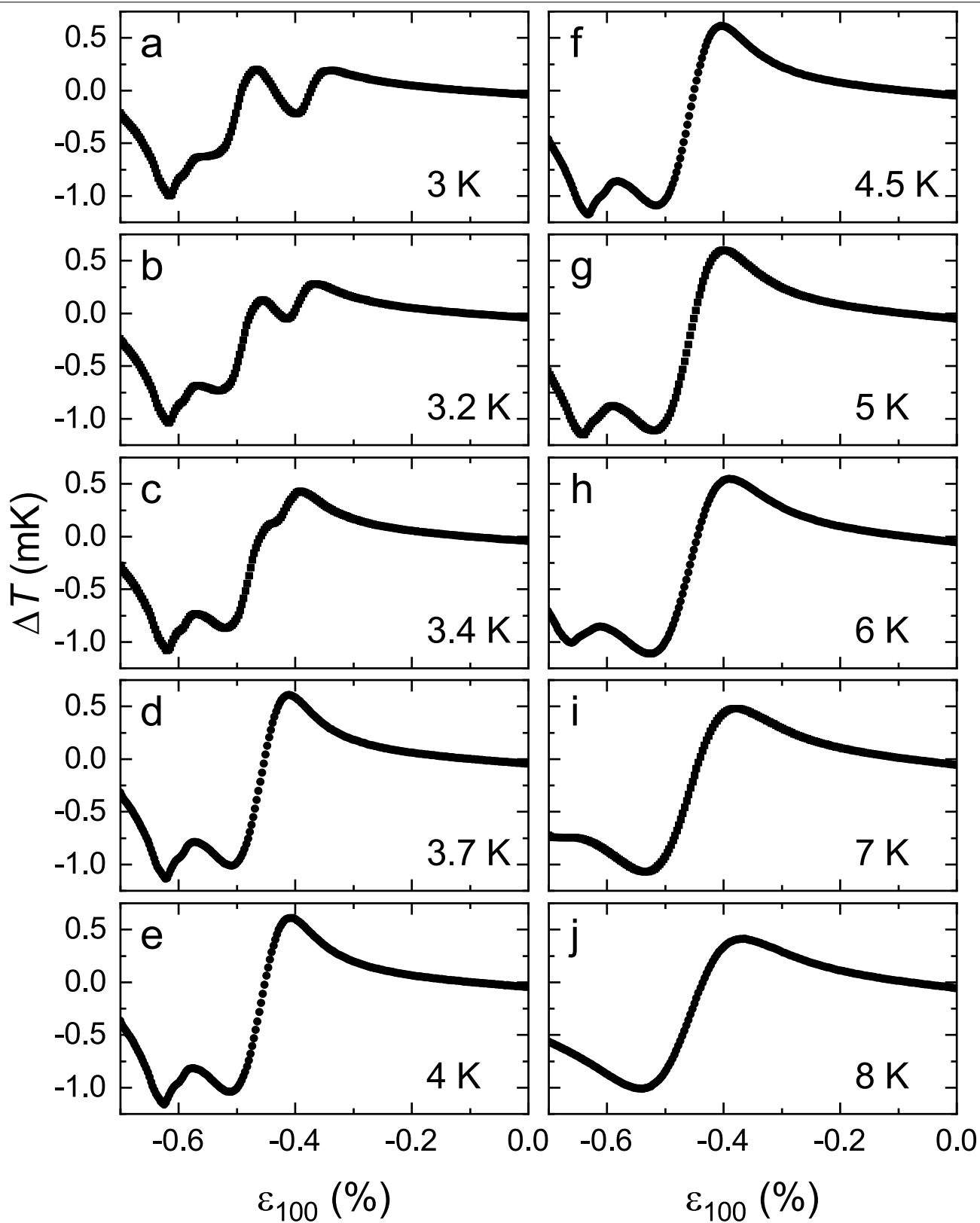

**Extended Data Fig. 2 | Response of the elastocaloric effect as a function of strain (further data: high temperatures).** $\Delta T(\varepsilon_{100})$ recorded at 1,513 Hz with an excitation amplitude $\varepsilon_{100}^{exc}$ between $2.9 \times 10^{-6}$ and $3.5 \times 10^{-6}$ during strain sweeps at ten different temperatures from 3 K to 8 K.

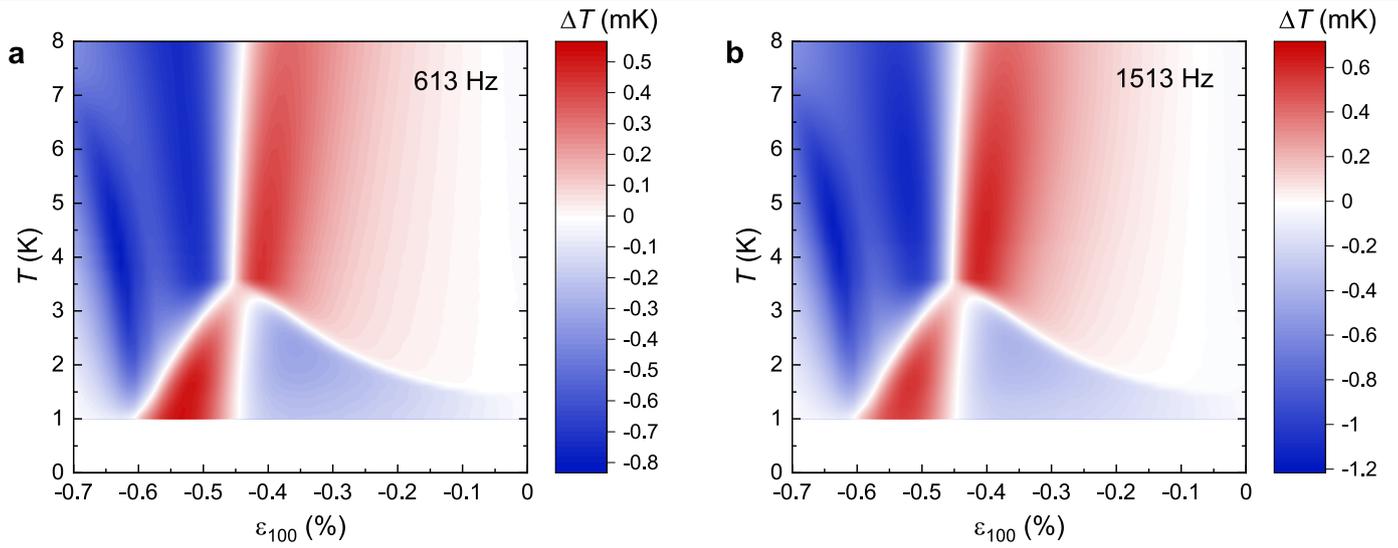

**Extended Data Fig. 3 | Comparison of the colour maps of the elastocaloric effect taken at different frequencies.** Data taken at 613 Hz (**a**) and at 1,513 Hz (**b**). The data were recorded with an excitation amplitude $\varepsilon_{100}^{exc}$ between $2.9 \times 10^{-6}$ and $3.5 \times 10^{-6}$.

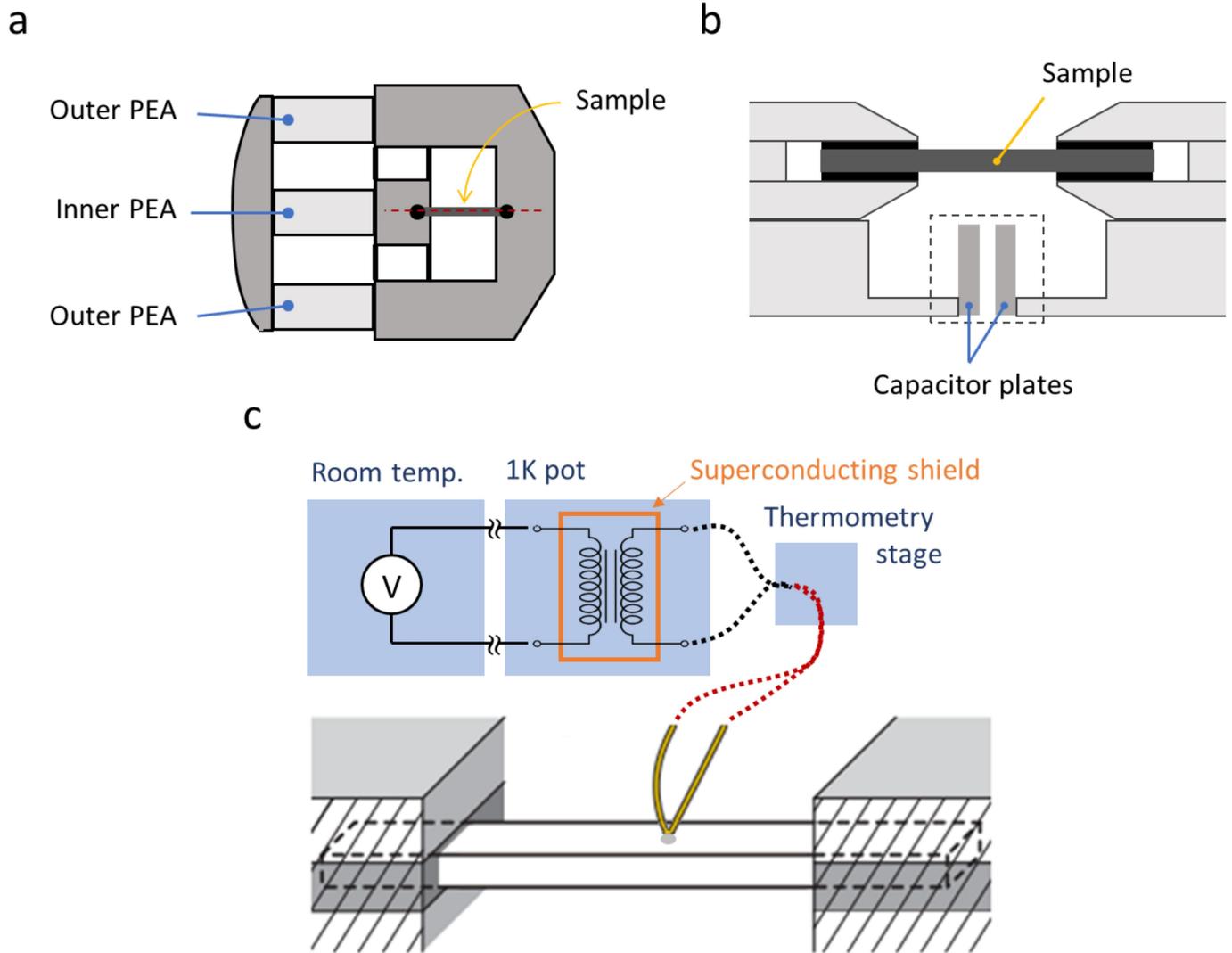

**Extended Data Fig. 4 | Schematic diagrams of the experimental setup.
a**, A uniaxial strain cell with inner and outer piezoelectric actuators (PEAs). To vary the tuning strain on the sample, the outer PEAs were driven between −350 V and 250 V. The inner PEA was kept constant at 185 V and the amplitude of the a.c. voltage to study the elastocaloric effect was between 0.25 V and 0.5 V. **b**, A detailed view of the mounted sample and the shielded capacitor plates, indicated by the dashed black box. The applied strain was determined by measuring the displacement of a capacitor mounted in parallel to the sample. **c**, The configuration of the electronic setup for the ultra-low-noise measurement of the temperature oscillations. The solid yellow lines represent the thermocouple, the dotted red lines the twisted copper wires and the dotted black lines twisted NbTi wires. The thermocouple wires are soldered to twisted copper wires, which were thermally anchored on the thermometry stage. From there, twisted superconducting wires, to reduce the input impedance, are connected to the input of a shielded low-temperature transformer, which is thermally anchored on the 1-K pot of the dilution refrigerator. Finally, the voltage is measured at room temperature using a lock-in amplifier.

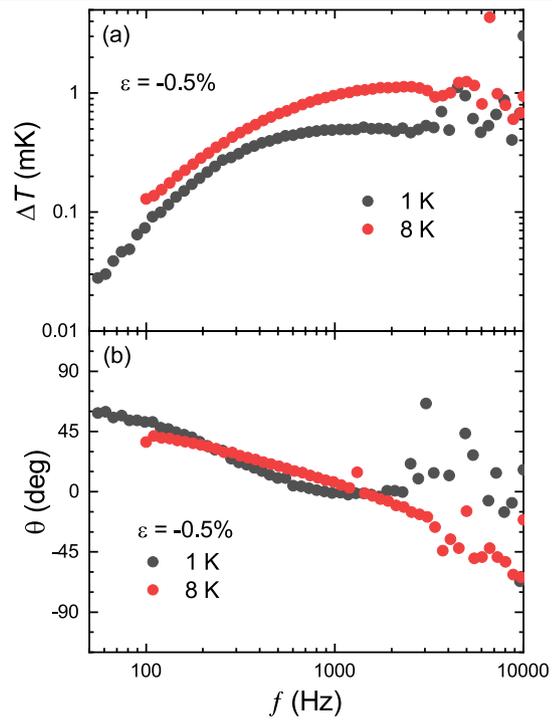

**Extended Data Fig. 5 | Frequency response of the thermocouple.**
Elastocaloric effect under 0.5% compression at 1 K and 8 K plotted against a logarithmic frequency scale. The applied strain oscillation is $\Delta\varepsilon = 3.45 \times 10^{-6}$.

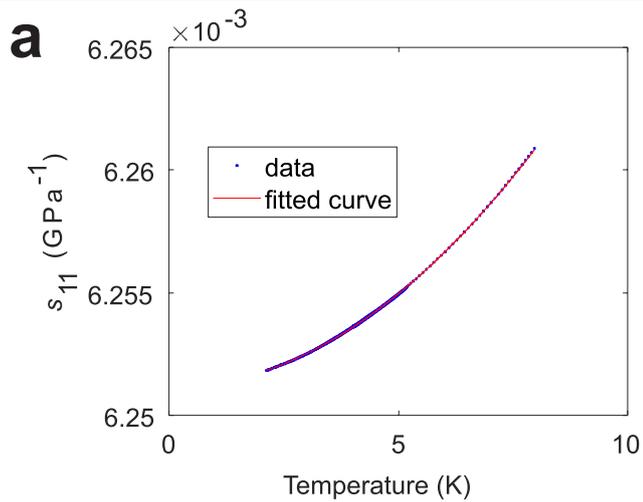 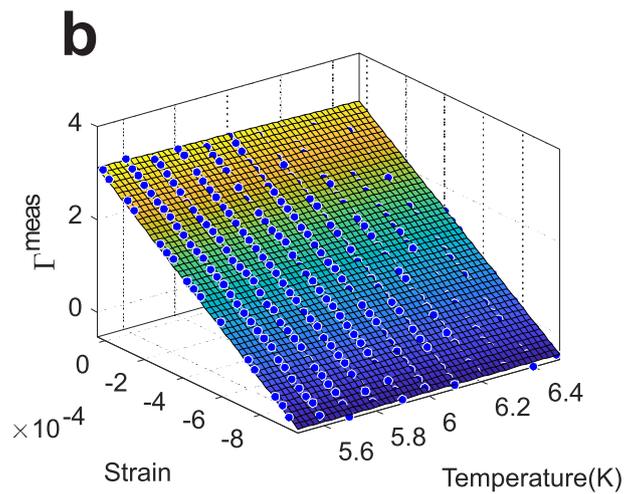

**Extended Data Fig. 6 | Data from a resonant ultrasound experiment and result of the calibration procedure of the Grüneisen parameter. a**, $s_{11}$ as a function of temperature determined in independent resonant ultrasound measurements together with a fit of the form shown in Equation (15). **b**, Measured Grüneisen parameter compared with the calibration function. The spheres correspond to the data and the surface to Equation (16) with the parameters mentioned in the text.

# Supplementary information

# Elastocaloric determination of the phase diagram of $Sr_2RuO_4$

# Elastocaloric determination of the phase diagram of $Sr_2RuO_4$

You-Sheng Li,[1] Markus Garst,[2,3] Jörg Schmalian,[3,4] Sayak Ghosh,[5] Naoki Kikugawa,[6] Dmitry A. Sokolov,[1] Clifford W. Hicks,[1,7] Fabian Jerzembeck,[1] Matthias S. Ikeda,[8,9,10] Zhenhai Hu,[1] B. J. Ramshaw,[5] Andreas W. Rost,[11,12] Michael Nicklas,[1] and Andrew P. Mackenzie[1,11]

[1]*Max Planck Institute for Chemical Physics of Solids,*
*Nöthnitzer Str. 40, 01187 Dresden, Germany*

[2]*Institut für Theoretische Festkörperphysik,*
*Karlsruher Institut für Technologie, 76131 Karlsruhe, Germany*

[3]*Institut für Quantenmaterialien und -technologien,*
*Karlsruher Institut für Technologie, 76131 Karlsruhe, Germany*

[4]*Institut für Theorie der Kondensierten Materie,*
*Karlsruher Institut für Technologie, 76131 Karlsruhe, Germany*

[5]*Laboratory of Atomic and Solid State Physics,*
*Cornell University, Ithaca, NY, USA*

[6]*National Institute for Materials Science, Tsukuba 305-0003, Japan*

[7]*School of Physics and Astronomy, University of Birmingham,*
*Birmingham B15 2TT, United Kingdom*

[8]*Geballe Laboratory for Advanced Materials, Stanford University,*
*476 Lomita Mall, Stanford, California 94305, USA*

[9]*Department of Applied Physics, Stanford University,*
*348 Via Pueblo Mall, Stanford, California 94305, USA*

[10]*Stanford Institute for Materials and Energy Science,*
*SLAC National Accelerator Laboratory,*
*2575 Sand Hill Road, Menlo Park, California 94025, USA*

[11]*Scottish Universities Physics Alliance, School of Physics and Astronomy,*
*University of St Andrews, St Andrews, UK*

[12]*Max Planck Institute for Solid State Research,*
*Heisenbergstr. 1, 70569 Stuttgart, Germany*

# I. FREQUENCY, TEMPERATURE, AND STRAIN DEPENDENCE OF THE AC STRAIN AMPLITUDE

In our study of the elastocaloric effect we used the *in situ* determined displacement amplitude $d_{exc}$ to calculate the applied AC strain amplitude $\Delta\varepsilon$. In the following we demonstrate that the applied displacement amplitude $d_{exc}$ displays only small variations as function of frequency, temperature, and strain in the parameter range of our measurements.

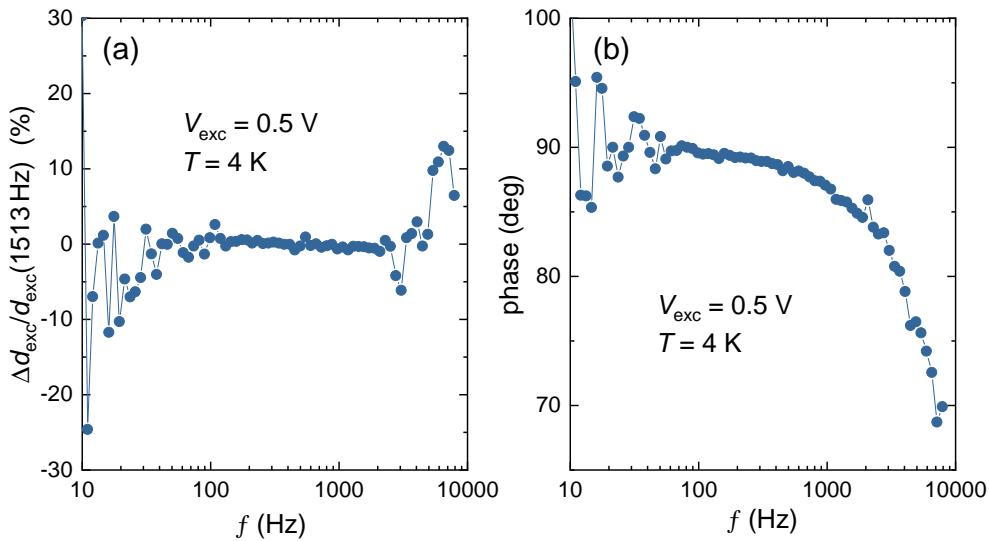

Figure S1. Frequency sweeps of (a) $d_{exc}$ and the corresponding (b) phase shift at zero applied tuning strain.

Figure S1 displays the frequency response of the AC displacement amplitude and the corresponding phase shift at 4 K for an excitation amplitude of 0.5 V. $d_{exc}$ is in a wide range independent of frequency. Below about 70 Hz we observe an increased scattering of the data. At high-frequency, above 2 kHz, we find an abrupt increase in $d_{exc}(f)$ indicating a nonlinear voltage-displacement response of the piezo-electric actuator. In this frequency range also the phase shift starts to deviate considerably from 90°. We conclude that in our set-up, the response of the actuators is appropriate for measurement in the range between 70 Hz and approximately 2 kHz.

The AC displacement amplitude exhibits a small temperature dependence. It monotonically increases by approximately 6% between 1 and 8 K. Figure S2 shows data taken at a frequency of 1513 Hz with an AC voltage amplitude $V_{exc} = 0.5$ V used to drive the piezo-electric actuator.

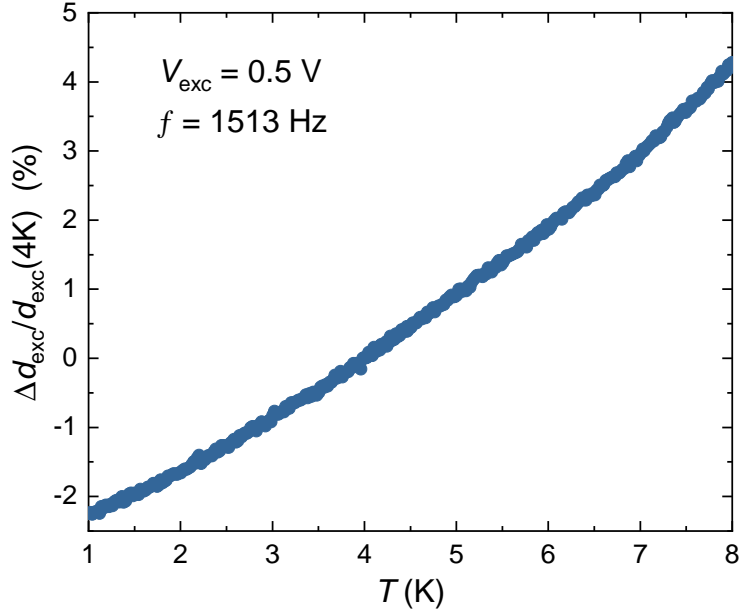

Figure S2. Temperature sweeps of $\Delta d_{exc} = [d_{exc} - d_{exc}(4\text{K})]/d_{exc}(4\text{K})$ at zero applied tuning strain.

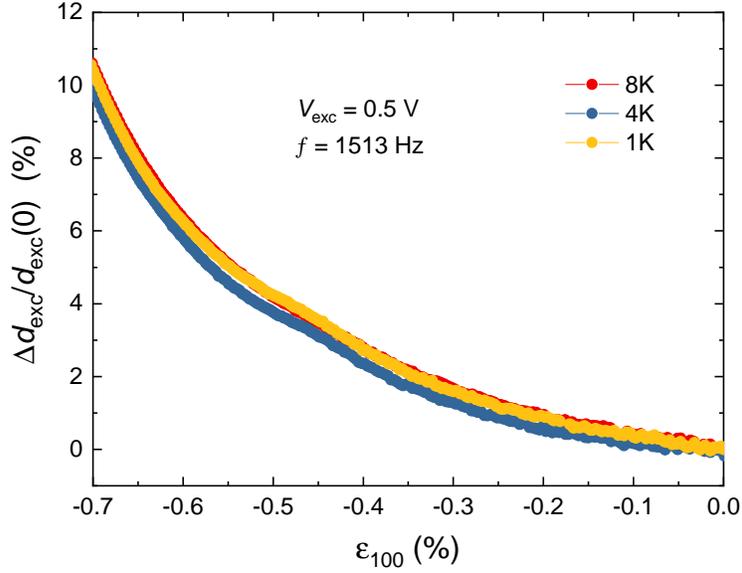

Figure S3. Tuning strain sweeps of $\Delta d_{exc} = [d_{exc} - d_{exc}(\varepsilon_{100} = 0)]/d_{exc}(\varepsilon_{100} = 0)$ at different temperatures.

The same frequency and AC voltage amplitude was used to investigate the dependence of $d_{exc}$ on applied tuning strain. Figure S3 displays data at three selected temperatures. The general behavior of $d_{exc}(\varepsilon)$ does not change with temperature. At all temperature the maximum change in $d_{exc}(\varepsilon)$ is less than 11 % over the range of strain used in our experiments.

## II. THEORETICAL CONSIDERATIONS

In section II A we provide a review on the relation between the elastocaloric effect and the uniaxial Grüneisen parameter $\Gamma$. In section II B we provide details of the calculation of $\Gamma$ across the superconducting transition close to the strain-induced Van-Hove singularity of $Sr_2RuO_4$ using a parametrisation of its $\gamma$-band; some results of this calculation are shown in the main text. Finally, in section II C we discuss the entropy quench across the superconducting transition within an effective theory valid close to the Van-Hove singularity, and we compare the scenario of a full gap versus a gap with nodes at the Van-Hove point in momentum space. We find that a nodal gap at the Van-Hove point is inconsistent with experimental signatures.

### A. Elastocaloric effect and uniaxial Grüneisen parameter

In the case that the temperature $T$ and uniaxial strain, say, along the $x$-axis $\varepsilon_{xx}$ can be controlled experimentally the entropy should be considered as a function $S = S(T, \varepsilon_{xx})$. Its differential is then given by

$$dS(T, \varepsilon_{xx}) = \left.\frac{\partial S}{\partial T}\right|_{\varepsilon_{xx}} dT + \left.\frac{\partial S}{\partial \varepsilon_{xx}}\right|_T d\varepsilon_{xx} = \frac{C_{\varepsilon_{xx}}}{T} dT + \left.\frac{\partial S}{\partial \varepsilon_{xx}}\right|_T d\varepsilon_{xx}, \tag{S1}$$

where $C_{\varepsilon_{xx}} = T\left.\frac{\partial S}{\partial T}\right|_{\varepsilon_{xx}}$ is the specific heat at constant uniaxial strain $\varepsilon_{xx}$. Under adiabatic conditions $dS = 0$, it follows

$$\Gamma = -\frac{1}{T}\left.\frac{dT}{d\varepsilon_{xx}}\right|_S = \frac{\left.\frac{\partial S}{\partial \varepsilon_{xx}}\right|_T}{C_{\varepsilon_{xx}}}, \tag{S2}$$

where $\Gamma$ is the Grüneisen parameter generalized to the case of uniaxial strain $\varepsilon_{xx}$. Up to a factor of $T$, it quantifies the change of temperature upon adiabatically varying $\varepsilon_{xx}$ and thus describes an elastocaloric effect. The Grüneisen parameter defined in Eq. S2 is dimensionless.

The derivative $\left.\frac{\partial S}{\partial \varepsilon_{xx}}\right|_T$ can be related to the uniaxial thermal expansion, $\alpha_{xx} = \left.\frac{\partial \varepsilon_{xx}}{\partial T}\right|_{\sigma_{zz}} = \left.\frac{\partial S}{\partial \sigma_{xx}}\right|_T$, and Young's modulus, $\frac{1}{E_Y} = \left.\frac{\partial \sigma_{xx}}{\partial \varepsilon_{xx}}\right|_T$,

$$\left.\frac{\partial S}{\partial \varepsilon_{xx}}\right|_T = \left.\frac{\partial S}{\partial \sigma_{xx}}\right|_T \left.\frac{\partial \sigma_{xx}}{\partial \varepsilon_{xx}}\right|_T = \frac{\alpha_{xx}}{E_Y}, \tag{S3}$$

where $\sigma_{xx}$ is the corresponding stress component. The uniaxial Grüneisen parameter is then given by $\Gamma = \alpha_{xx}/(C_{\varepsilon_{xx}} E_Y)$.

## B. Grüneisen parameter arising from the $\gamma$-band of $Sr_2RuO_4$

### 1. Tight-binding model for the $\gamma$-band

We follow Refs.[1, 2] and consider an experimentally determined tight-binding model of the $\gamma$-band of $Sr_2RuO_4$ with the two-dimensional electron spectrum

$$\varepsilon_{\bm{k}} = -2t_x \cos(k_x a_x) - 2t_y \cos(k_y a_y) - 4t' \cos(k_x a_x) \cos(k_y a_y) - \mu \quad (S4)$$

where $\mu$ is the chemical potential. The nearest-neighbour hopping amplitudes are $t_x$ and $t_y$, and $t'$ is the next-nearest-neighbour hopping amplitude within the crystallographic $(a, b)$-plane with lattice constants $a_x$ and $a_y$.

In the absence of strain the system possesses tetragonal symmetry with lattice constants $a_x = a_y$. In order to obtain the values for the hopping parameters at zero strain, we make use of tight-binding fits to the ARPES data by Burganov et al. [3] that yields the values $t_x = t_y = t_0 = 0.119$ eV, $t'_0 = 0.392 t_0$, and $\mu = 1.48 t_0$. The value for $t_0$ and the chemical potential $\mu$ are directly taken from Ref. [3]. For the next-nearest neighbour hopping Burganov et al. give the value $t'_0 = 0.41 t_0$; however, this value yields an energy of 19 meV at the point $(k_x, k_y) = (0, \pi)$ in the two-dimensional Brillouin zone that overestimates the value obtained experimentally. Instead of a global fit performed by Burganov et al. [3], we refitted their data with a local fit optimizing the behavior around this point in the Brillouin zone yielding a value for $t'_0$ that is 4% smaller. We note that the value for $t'_0$ of Burganov et al. [3] will not qualitatively change the behavior of the Grüneisen parameter but change its magnitude by approximately a factor of two due to an overestimate of the electronic energy.

In the presence of normal strains $\varepsilon_{xx}$ and $\varepsilon_{yy}$, the amplitudes are modified, and for small strains we can approximate up to linear order

$$t_x = t_0(1 - \alpha \varepsilon_{xx}), \quad t_y = t_0(1 - \alpha \varepsilon_{yy}), \quad t' = t'_0(1 - \alpha'(\varepsilon_{xx} + \varepsilon_{yy})/2), \quad (S5)$$

with linear coefficients $\alpha$ and $\alpha'$. The uniaxial strain along the $y$-direction, $\varepsilon_{yy} = -\nu_{xy} \varepsilon_{xx}$, is determined by $\varepsilon_{xx}$ with the Poisson ratio whose value at ambient pressure and 4 K is given by $\nu_{xy} \approx 0.508$ [4]. The Van-Hove singularity is reached in the experiment for a strain of $\varepsilon_{xx} = -0.44\%$, see above. The tight-binding dispersion exhibits a Van-Hove point at $(k_x, k_y) = (0, \pi)$ for a choice of the parameters $\alpha = \alpha' \approx 7.604$, where we used $\alpha = \alpha'$ for simplicity. This value is consistent with the calculations of Barber *et al.* [5]. The parameter $\alpha$ will determine the overall magnitude of the Grüneisen parameter.

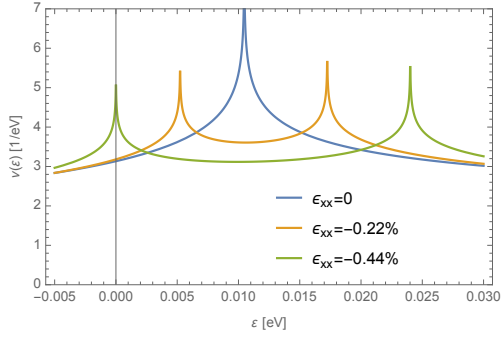

Figure S4. Density of states S6 associated with the $\gamma$-band of Sr$_2$RuO$_4$ in units of 1/eV as a function of energy for various values of uniaxial strain $\varepsilon_{xx}$. A single Van-Hove singularity for zero strain splits into two for finite $\varepsilon_{xx}$. A Van-Hove singularity reaches the Fermi level at $\varepsilon = 0$ for $\varepsilon_{xx} = -0.44\%$ (green line).

The corresponding density of states per spin is given by

$$\nu(\varepsilon) = a_x a_y \int_{-\pi/a_x}^{\pi/a_x} \frac{dk_x}{2\pi} \int_{-\pi/a_y}^{\pi/a_y} \frac{dk_y}{2\pi} \delta(\varepsilon - \varepsilon_{\bm{k}}). \tag{S6}$$

We have multiplied the integral by the area of the two-dimensional unit cell, $a_x a_y$, so that $\nu(\varepsilon)$ represent the number of states per energy and per unit cell. This density of states is independent of the lattice constants, as becomes manifest upon substituting the integration variables $k_x = \tilde{k}_x/a_x$ and $k_y = \tilde{k}_y/a_y$. The strain dependence of $\nu(\varepsilon)$ thus arises from the hopping amplitudes in the dispersion of Eq. S4. $\nu(\epsilon)$ and other densities of states that we use in the analysis of the Grüneisen parameter can be expressed in terms of elliptic integrals.

The evolution of the energy-dependent density of states with uniaxial strain is illustrated in Fig. S4. The Van-Hove singularity at zero uniaxial strain splits into two for finite $\varepsilon_{xx}$. A Van-Hove singularity reaches the Fermi level for $\varepsilon_{xx} = -0.44\%$. The density of states at the Fermi level $\nu(0)$ as a function of uniaxial strain is shown in Fig. S5. A Van-Hove singularity at the Fermi level is realized for compressive strain $\varepsilon_{xx} \approx -0.44\%$ as well as tensile strain $\varepsilon_{xx} \approx 0.34\%$. In between the density of states is minimal for $\varepsilon_{xx} \approx -0.07\%$.

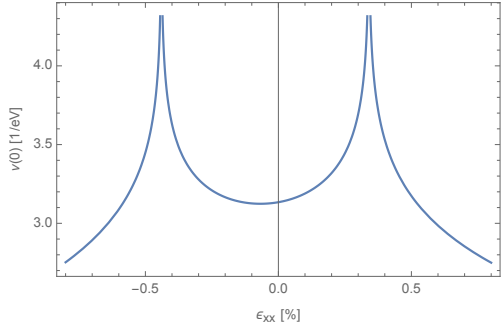

Figure S5. Density of states S6 at the Fermi level $\nu(0)$ as a function of uniaxial strain $\varepsilon_{xx}$. A Van-Hove singularity at the Fermi level is realized for compressive strain $\varepsilon_{xx} \approx -0.44\%$ as well as tensile strain $\varepsilon_{xx} \approx 0.34\%$. As a result, a minimum arises at $\varepsilon_{xx} \approx -0.07\%$.

2. *Thermodynamics arising from the $\gamma$-band*

In order to account for the behavior of the Grüneisen parameter across the superconducting transition close to the Van-Hove singularity of $Sr_2RuO_4$, we consider the entropy per unit cell,

$$S(T, \varepsilon_{xx}) = -2k_B \int d\varepsilon\, \nu(\varepsilon)\Big(f(E)\log f(E) + (1 - f(E))\log(1 - f(E))\Big) \quad (S7)$$

where $E(\varepsilon) = \sqrt{\varepsilon^2 + \Delta^2}$ with the superconducting gap $\Delta$, and the Fermi function $f(E) = (e^{E/(k_B T)} + 1)^{-1}$. The factor of 2 in front of the integral of Eq. S7 accounts for the spin degree of freedom. The superconducting gap $\Delta$ is determined by the gap equation

$$\frac{1}{g} = \int_{-\hbar\omega_D}^{\hbar\omega_0} d\varepsilon\, \nu(\varepsilon) \frac{\tanh \frac{E}{2k_B T}}{2E} \quad (S8)$$

with a cutoff provided by the frequency $\omega_0$, and the coupling constant $g > 0$ with units of energy. Our choice for the cutoff frequency $\hbar\omega_0/k_B = 400$ K was informed by inelastic neutron scattering data on $Sr_2RuO_4$ [6]. In order to reproduce the critical temperature $T_c \approx 3.5$ K as observed at the strain-induced Van-Hove singularity of $Sr_2RuO_4$, we need to choose $g/k_B \approx 714$ K. The gap $\Delta = \Delta(T, \varepsilon_{xx})$ itself will not only depend on temperature $T$ but also on strain $\varepsilon_{xx}$, see Figure S6.

With the help of the entropy S7 the Grüneisen parameter can be evaluated using Eq. S2 by taking derivatives with respect to $T$ and strain $\varepsilon_{xx}$. The specific heat is given by

$$C_{\varepsilon_{xx}} = T\frac{\partial S}{\partial T}\bigg|_{\varepsilon_{xx}} = \frac{2k_B}{4(k_B T)^2} \int d\varepsilon\, \nu(\varepsilon) \frac{E^2 - T\Delta\partial_T\Delta}{\cosh^2(E/(2k_B T))}. \quad (S9)$$

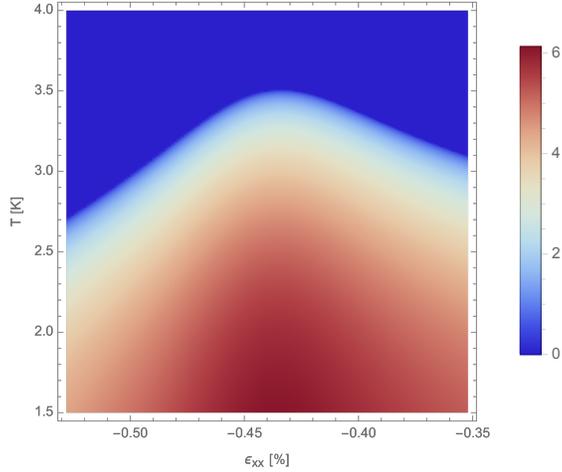

Figure S6. Superconducting gap $\Delta/k_B$ evaluated with Eq. S8 in units of Kelvin as a density plot within the $(\varepsilon_{xx}, T)$ plane. The maximal critical temperature $T_c|_{\max} \approx 3.5$ K is obtained close to the Van-Hove singularity $\varepsilon_{xx} \approx -0.44\%$.

For the strain derivative we obtain

$$\left.\frac{\partial S}{\partial \varepsilon_{xx}}\right|_T = \frac{2k_B}{4(k_BT)^2} \int d\varepsilon \, \frac{\nu_{\varepsilon_{xx}}(\varepsilon)\varepsilon - \nu(\varepsilon)\Delta \partial_{\varepsilon_{xx}}\Delta}{\cosh^2(E/(2k_BT))}, \tag{S10}$$

where we introduced the auxiliary function

$$\nu_{\varepsilon_{xx}}(\varepsilon) = a_x a_y \int_{-\pi/a_x}^{\pi/a_x} \frac{dk_x}{2\pi} \int_{-\pi/a_y}^{\pi/a_y} \frac{dk_y}{2\pi} \frac{\partial \varepsilon_{\boldsymbol{k}}}{\partial \varepsilon_{xx}} \delta(\varepsilon - \varepsilon_{\boldsymbol{k}}). \tag{S11}$$

In the normal phase $\Delta = 0$ sufficiently far away from the Van Hove singularity where the density of states $\nu(\varepsilon)$ only smoothly varies in the range $|\varepsilon| \lesssim k_B T$ around the Fermi energy, standard Fermi liquid behaviour is obtained. The entropy then depends linearly on temperature

$$S \approx \frac{2\pi^2}{3} k_B^2 T \nu(0), \tag{S12}$$

and the Grüneisen parameter reduces to a constant

$$\Gamma \approx \frac{1}{\nu(0)} \frac{\partial \nu(0)}{\partial \varepsilon_{xx}}, \tag{S13}$$

that quantifies the strain dependence of the density of states at the Fermi level $\nu(0)$. From this expression follows that the Grüneisen parameter in the Fermi-liquid limit will change sign at maxima as well as minima of $\nu(0)$ as a function of uniaxial strain $\varepsilon_{xx}$. The pronounced sign change of $\Gamma$ close to the Van Hove singularity $\varepsilon_{xx} = -0.44\%$ where $\nu(0)$ is maximal is clearly observed in the experimental data, see figure 3 in the main text. The additional sign change in the experimental data close to zero strain is in agreement with the expected minimum of $\nu(0)$, see Fig. S5.

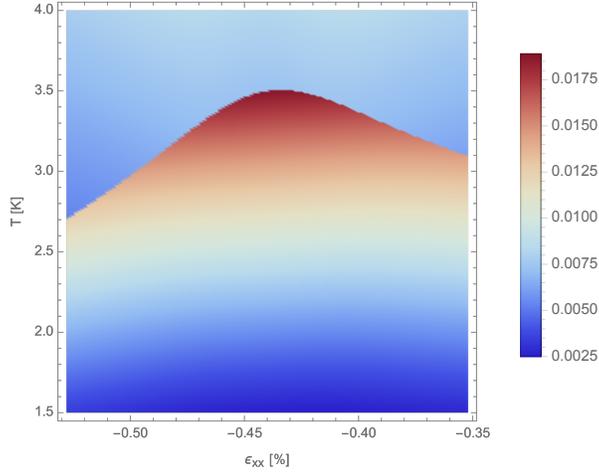

Figure S7. Specific heat $C_{\varepsilon_{xx}}$ evaluated with Eq. S9 in units of $k_B$ per unit cell as a density plot within the $(\varepsilon_{xx}, T)$ plane.

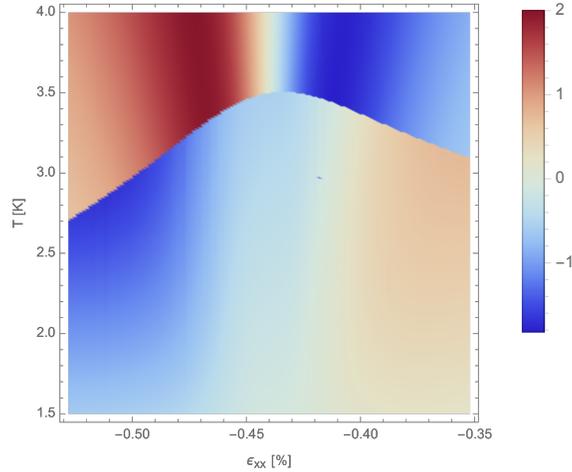

Figure S8. Strain derivative of the entropy $\partial S/\partial \varepsilon_{xx}|_T$ evaluated with Eq. S10 in units of $k_B$ per unit cell as a density plot within the $(\varepsilon_{xx}, T)$ plane.

The evaluation of the specific heat, the strain derivative of entropy and the Grüneisen parameter are shown as a function of temperature $T$ and uniaxial strain $\varepsilon_{xx}$ in Figs. S7, S8 and S9, respectively. Additional line cuts are shown in the main text. At the Van Hove singularity above the critical temperature, entropy accumulates resulting in a maximum of $S(\varepsilon_{xx})$ and thus a sign change of $\partial S/\partial \varepsilon_{xx}|_T$ and $\Gamma$. As the superconducting phase is entered upon lowering the temperature, a large part of the entropy is quenched so that the maximum in $S(\varepsilon_{xx})$ is converted into a minimum. This leads to a saddle point in the strain dependence of the entropy $S(\varepsilon_{xx})$ close to the maximum of the critical temperature $T_c$.

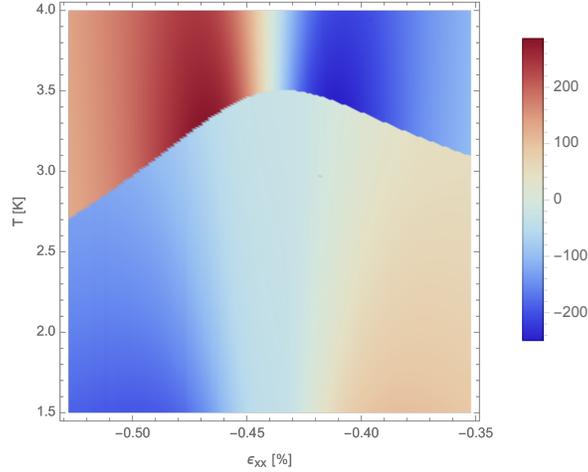

Figure S9. Grüneisen parameter $\Gamma = (\partial S/\partial \varepsilon_{xx}|_T)/C_{\varepsilon_{xx}}$ using the results of Figs. S7 and S8 as a density plot within the $(\varepsilon_{xx}, T)$ plane.

### 3. Influence of inter-layer hopping on the Van Hove singularity

A finite inter-plane hopping of electrons will eventually cut off the logarithmic divergence of the Van Hove singularity in the density of states and, as a consequence, weaken and broaden the associated thermodynamic signatures. Here, we estimate the temperature and strain scale at which this cut-off is expected to happen.

We start the discussion with an extended tight-binding Hamiltonian following Ref. [7] that is based on the three $t_{2g}$ Ru orbitals,

$$\mathcal{H}_s(\vec{k}) = \begin{pmatrix} \varepsilon_{AA}(\vec{k}) & \varepsilon_{AB}(\vec{k}) - is\eta & \varepsilon_{AC}(\vec{k}) + i\eta \\ \varepsilon_{AB}(\vec{k}) + is\eta & \varepsilon_{BB}(\vec{k}) & \varepsilon_{BC}(\vec{k}) - s\eta \\ \varepsilon_{AC}(\vec{k}) - i\eta & \varepsilon_{BC}(\vec{k}) - s\eta & \varepsilon_{CC}(\vec{k}) \end{pmatrix}. \tag{S14}$$

Here, $s = \pm 1$ represents the spin degree of freedom, $\eta$ is the spin-orbit coupling, and the dispersions are parameterised as

$$\varepsilon_{AA}(k_x, k_y, k_z) = -2t_1\cos(k_x) - 2t_2\cos(k_y) - 2t_3\cos(2k_x) - 4t_4\cos(k_x)\cos(k_y) \quad \text{(S15)}$$
$$-4t_5\cos(2k_x)\cos(k_y) - 2t_6\cos(3k_x) - 2t_7\cos(2k_y)$$
$$-2t_8\cos(k_x/2)\cos(k_y/2)\cos(k_z) - \mu_{1D},$$
$$\varepsilon_{BB}(k_x, k_y, k_z) = \varepsilon_{AA}(k_y, k_x, k_z), \quad \text{(S16)}$$
$$\varepsilon_{CC}(k_x, k_y, k_z) = -2\bar{t}_1(1-\alpha\varepsilon_{xx})\cos(k_x) - 2\bar{t}_1(1+\alpha\nu_{xy}\varepsilon_{xx})\cos(k_y) \quad \text{(S17)}$$
$$-4\bar{t}_2(1-\frac{\alpha'}{2}(1-\nu_{xy})\varepsilon_{xx})\cos(k_x)\cos(k_y) - 2\bar{t}_3(\cos(2k_x)+\cos(2k_y))$$
$$-4\bar{t}_4(\cos(2k_x)\cos(k_y)+\cos(2k_y)\cos(k_x))$$
$$-\bar{t}_5\cos(k_x/2)\cos(k_y/2)\cos(k_z) - \mu_{2D},$$
$$\varepsilon_{AB}(k_x, k_y, k_z) = -4t_{\text{int},1}\sin(k_x)\sin(k_y) - 4t_{\text{int},2}\sin(k_x/2)\sin(k_y/2)\cos(k_z), \quad \text{(S18)}$$
$$\varepsilon_{AC}(k_x, k_y, k_z) = -4t_{\text{int},3}\cos(k_x/2)\sin(k_y/2)\sin(k_z), \quad \text{(S19)}$$
$$\varepsilon_{BC}(k_x, k_y, k_z) = -4t_{\text{int},3}\cos(k_y/2)\sin(k_x/2)\sin(k_z). \quad \text{(S20)}$$

We used already dimensionless wavevectors, $a_i k_i \to k_i$ with $i = x, y, z$ and $k_i \in (-\pi, \pi]$. This Hamiltonian gives rise to the $\alpha$, $\beta$, and $\gamma$ bands. The values of the hopping parameters can be found in Ref. [7] where they were obtained by a fit to *ab initio* calculations. We already accounted for a strain dependence $\varepsilon_{xx}$ of the hopping parameters $\bar{t}_1$ and $\bar{t}_2$ with a strength $\alpha$ and $\alpha'$, similar to Eq. S5, and neglected for simplicity the dependence of all other hopping parameters on strain.

Note that the *ab initio* parameters of Ref. [7] overestimate the values of the hopping parameters. For example, the value $\bar{t}_1 = 0.3568$ eV [7] should be compared to $t_0 = 0.119$ eV of section II B 1. Using the *ab initio* hopping parameters of Ref. [7] and the Poisson ratio $\nu_{xy} = 0.508$ [4] setting $\alpha' = \alpha$ for simplicity, the $\gamma$ band dispersion $\varepsilon_\gamma(\vec{k})$ of Eq. S14 vanishes at $(k_x, k_y, k_z) = (0, \pi, 0)$ for the critical strain $\varepsilon_{xx}^{\text{VH}} = -0.44\%$ for a value of the parameter $\alpha = 15.62$. These values based on *ab initio* calculations will be used for the following estimates.

First, let us recall the behavior of the $\gamma$ band of the two-dimensional model of section II B 1. Close to the point $(k_x, k_y) = (0, \pi)$ its dispersion possesses a saddle point

$$\varepsilon_\gamma(k_x, k_y, 0) \approx c_1 k_x^2 - c_2(k_y - \pi)^2 + c_3\delta\varepsilon_{xx}, \quad \text{(S21)}$$

with positive coefficients $c_1, c_2, c_3$, and $\delta\varepsilon_{xx} = \varepsilon_{xx} - \varepsilon_{xx}^{\rm VH}$ is the distance to the critical strain. This leads to a logarithmic singularity in the two-dimensional density of states

$$\nu_{2d}(\varepsilon) = \int_{-\pi}^{\pi} \frac{dk_x}{2\pi} \int_{-\pi}^{\pi} \frac{dk_y}{2\pi} \delta(\varepsilon - \varepsilon_\gamma(k_x, k_y, 0)) \sim \frac{1}{8\pi^2\sqrt{c_1 c_2}} \log \frac{1}{|\varepsilon - c_3 \delta\varepsilon_{xx}|}. \tag{S22}$$

The dispersion of the $\gamma$ band resulting from the full three-dimensional model of Eq. S14 possesses close to $(k_x, k_y, 0) = (0, \pi, k_z)$ the modified form

$$\varepsilon_\gamma(\vec{k}) \approx c_1 k_x^2 - c_2(k_y - \pi)^2 + c_3 \delta\varepsilon_{xx} + c_4(k_y - \pi)\cos(k_z) - c_5 k_x(k_y - \pi)\sin(k_z) + c_6 \sin^2(k_z). \tag{S23}$$

We kept here only the first order Fourier components of $k_z$ except for the last term with coefficient $c_6$ that involves a second order $2k_z$ Fourier component. We also neglected contributions of order $\mathcal{O}(k_x^2 \delta\varepsilon)$ and $\mathcal{O}((k_y - \pi)^2 \delta\varepsilon)$. Using the values of the hopping parameters of Ref. [7] we obtain for the coefficients the following estimates $c_2 \approx 0.395$ eV, $c_1/c_2 \approx 0.15$, $c_4/c_2, c_5/c_2 \sim 0.01$, $c_6 \approx 0.0005$ eV and $c_3 \approx 12.950$ eV. The contribution $c_4$ and $c_5$ shift the position of the saddle point and can effectively be absorbed into small shift of $c_1, c_2$ and $c_6$ by substitution of the wavevectors $(k_x, k_y)$. The main effect arises from the coefficient $c_6$ that effectively leads to a $k_z$-dependence of the tuning parameter $\delta\varepsilon_{xx}$. It cuts off the logarithmic singularity in the three dimensional density of states,

$$\nu_{3d}(\varepsilon) \sim \frac{1}{8\pi^2\sqrt{c_1 c_2}} \int_{-\pi}^{\pi} \frac{dk_z}{2\pi} \log \frac{1}{|\varepsilon - c_3 \delta\varepsilon_{xx} - c_6 \sin^2(k_z)|}. \tag{S24}$$

The resulting density of states for $\delta\varepsilon_{xx} = 0$ is approximately constant in the range $|\varepsilon| < c_6$ with two cusps, i.e., typical non-analyticities for a three-dimensional density of states at the two edges $\varepsilon = \pm c_6$, see Fig. S10(a).

The regularization of the logarithmic singularity in the density of states by the inter-layer hopping leads to a temperature scale $T_{3d} \sim c_6/k_B$ and a strain scale $\varepsilon_{xx}^{3d} \sim c_6/c_3$. Using the above estimates we obtain the values $T_{3d} \approx 5.8$ K and $\delta\varepsilon_{xx}^{3d} \sim 3.8 \times 10^{-5}$, i.e., approximately $10^{-2}$ of the absolute value of the critical strain. The latter is consistent with comparing $T_{3d}$ and the energy of the $\gamma$ band at $(k_x, k_y, k_z) = (0, \pi, 0)$ and zero strain that is 53 meV or 615 K using the *ab initio* parameters (compared to just over 10 meV measured experimentally). Both scales are reflected for example in the dependence of the specific heat coefficient of the normal state. In the low temperature limit, on the one hand, the specific heat coefficient is proportional to the density of states. In order to resolve the inter-layer regularization a strain resolution of order $\delta\varepsilon_{xx}^{3d}$ would be required that is far beyond the experimental resolution. On the other hand, at the critical strain the specific heat coefficient diverges logarithmically as a function of decreasing temperature for the two-dimensional model. For the three dimensional model with inter-layer hopping this divergence is cut off, and the specific heat coefficient saturates but only below a temperature of order $0.1 T_{3d} \sim 0.6$ K, see Fig. S10(b).

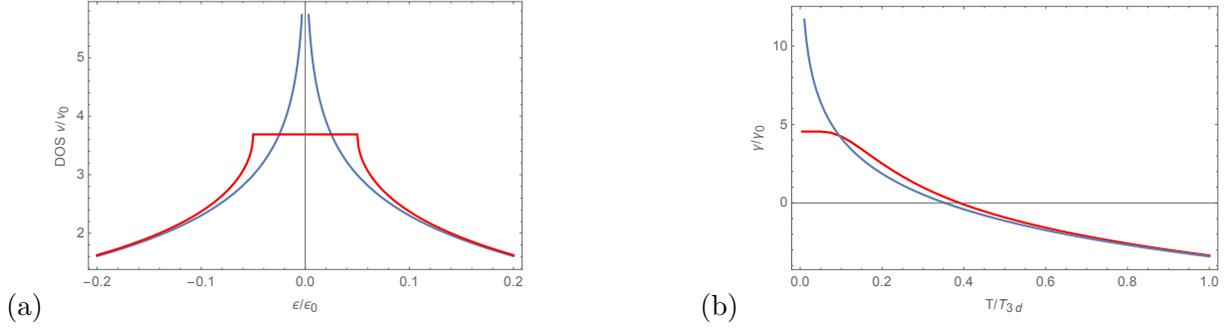

Figure S10. (a) Regularization of the two-dimensional Van Hove singularity in the density of states $\nu(\varepsilon) = \nu_0 \log(\varepsilon/\varepsilon_0)$ (blue line) at the critical strain $\delta\varepsilon_{xx} = 0$ via the inter-plane hopping that cuts off the singularity giving rise to two cusps (red line). For this illustration the coefficient was chosen to be $c_6 = 0.1\varepsilon_0$. (b) Specific heat coefficient $\gamma = C/T$ for the density of states of panel (a) (up to a constant background) as a function of temperature with $\gamma_0 = \nu_0 k_B$ where $k_B$ is the Boltzmann constant. The two-dimensional logarithmic divergence (blue line) is cut off by the inter-layer hopping and the specific heat coefficient saturates below a temperature of order $0.1 T_{3d}$.

We can conclude for $Sr_2RuO_4$ that its thermodynamics in the normal state is only barely affected by the inter-layer hopping. First, its signatures as a function of strain occur on a scale beyond the strain resolution of the experiment. Second, its signatures as a function of temperature are most pronounced at the critical strain but are negligible for temperatures $T > T_{c,\mathrm{max}} \approx 3.5$ K $\approx 0.6 T_{3d}$. For lower temperatures the Van Hove singularity seems to be rather cut off by the superconducting order parameter masking the effect of the inter-layer hopping on the Van Hove singularity in density of states. Finally, note that the *ab initio* parameters overestimate the scales so that our estimates about the influence of inter-layer hopping are on the conservative side.

### 4. Influence of disorder on the Van Hove singularity

The Van Hove singularity might in principle also be smeared out by elastic scattering off impurities. This becomes relevant once the elastic scattering rate $\hbar\tau_0^{-1}$ exceeds the scale for inter-layer coupling, which corresponds to several Kelvin [8]. We are, however, confident that the scattering rate must be significantly below this value. $Sr_2RuO_4$ is known to be sensitive to pair-breaking due to non-magnetic impurities [9]. Hence, the very observation of a superconducting transition implies that $\hbar\tau_0^{-1} \ll k_B T_c$ and elastic scattering events will not smear out the Van Hove singularity in the temperature regime of our measurements.

### C. Entropy quench for a superconducting gap with nodes at the Van Hove singularity

In section II B it was assumed that the superconductor is fully gapped at the Van Hove point in momentum space. In this case, the calculated entropy, that is enhanced at the Van Hove singularity, gets strongly quenched upon entering the superconducting phase, see Fig. S9. This behavior is consistent with the experimental observations as presented in the main text. In the present section, we demonstrate that a superconducting gap with nodes at the Van Hove points is not able to reproduce such a substantial entropy quench.

#### 1. Effective low-energy theory for the Van Hove singularity and nodal superconductivity

For this purpose, we consider an effective theory that is only valid close to the Van Hove singularity. Expanding the dispersion of Eq. S4 close to the Van Hove singularity at $k_y = \pi/a_y$, it acquires in lowest order the following form

$$\varepsilon_{\boldsymbol{q}} = \frac{1}{2m}(q_x^2 - q_y^2) + r \tag{S25}$$

where $r = -\mu + 2(t_y - t_x) + 4t'$ tunes the distance to the Van Hove point, the mass $m = \frac{1}{2a_x a_y \sqrt{(t_x - 2t')(t_y + 2t')}}$ and the wavevectors are $q_y = (k_y - \frac{\pi}{a_y})\sqrt{b}$ and $q_x = k_x/\sqrt{b}$ with the dimensionless scaling factor $b = \frac{a_y}{a_x}\sqrt{\frac{t_y + 2t'}{t_x - 2t'}}$. We will supplement the effective theory with a hard energy cutoff $(q_x^2 + q_y^2)/(2m) \leq \varepsilon_0$ where $\varepsilon_0$ is on the order of the hopping; we will use $\varepsilon_0 = \frac{1}{2ma_x a_y}$.

We assume that the superconducting gap depends in general analytically on momentum, $\Delta_{\bm{q}} = \Delta \gamma_{\bm{q}}$, and we consider the following three scenarios:

$$\gamma_{\bm{q}} = \begin{cases} 1 & \text{scenario (1)} \\ a_x a_y (q_x^2 - q_y^2) & \text{scenario (2)} \\ a_x a_y 2 q_x q_y & \text{scenario (3).} \end{cases} \tag{S26}$$

The scenario (1) corresponds to a hard gap at the Van Hove singularity, and this should reproduce the asymptotic behavior of section II B. The scenarios (2) and (3) correspond to a nodal superconductor. Whereas for case (2) the location of the nodes coincides with the location of the Fermi surface at $r = 0$, the scenario (3) still produces a gap for a generic point on the Fermi surface for $r = 0$ that decreases, however, and vanishes with distance to the location of the Van Hove point $|\bm{q}| = 0$ in momentum space.

The superconductor possesses the dispersion $E_{\bm{q}} = \sqrt{\varepsilon_{\bm{q}}^2 + \Delta_{\bm{q}}^2}$. For a momentum-dependent gap, it is convenient to consider in this section the density of states defined as follows

$$\nu(E) = a_x a_y \int_{\varepsilon_0} \frac{dq_x dq_y}{(2\pi)^2} \delta(E - \sqrt{\varepsilon_{\bm{q}}^2 + \Delta_{\bm{q}}^2}), \tag{S27}$$

where the integral over wavevectors is evaluated with the hard energy cutoff $\varepsilon_0 \geq (q_x^2 + q_y^2)/(2m)$. It is clear that $\nu(E)$ is symmetric in $r$ because the case $r < 0$ can be mapped onto $r > 0$ by interchanging $q_x$ and $q_y$ in the definition of $\nu(E)$. In the following, we therefore focus on positive $r > 0$.

The density of states is illustrated in Fig. S11 for the three gap scenarios of Eq. S26. The fully gapped scenario (1) produces a hard gap in the density of states $E_{\text{gap}} = \Delta$. The nodal scenario (2) also produces a small hard gap whose size however depends on both the tuning parameter and the cutoff, $E_{\text{gap}} = \frac{\Delta |r|}{\sqrt{\varepsilon_0^2 + \Delta^2}}$; right at the Van Hove point $r = 0$, the singularity in $\nu(E)$ is maintained. The nodal scenario (3) caps the Van Hove singularity in the density of states at $r = 0$ and gives rise to a plateau with constant $\nu(E)$ up to energies $E \leq \Delta$; for any finite $r \neq 0$ a soft gap appears.

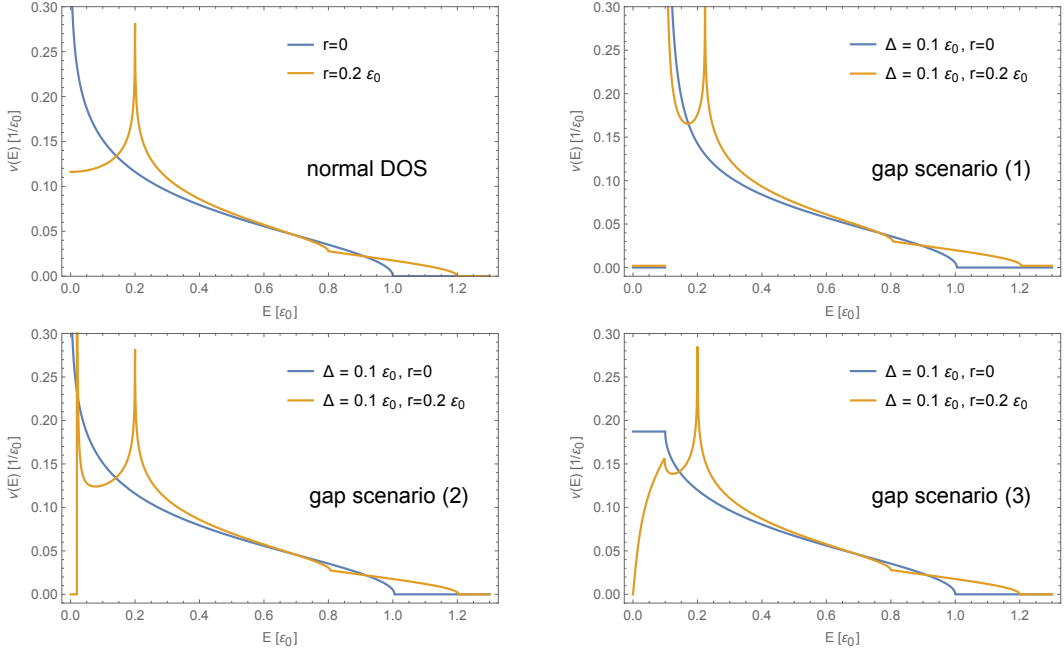

Figure S11. Density of states S27 for various values of the tuning parameter $r$ and the gap $\Delta$. The different panels illustrate the three scenarios for the superconducting gap, see Eq. S26.

2. *Gap equation and phase diagram*

The gap $\Delta$ fulfils the gap equation

$$\frac{1}{g} = \int_0^{E_0} dE\, \nu_\gamma(E) \frac{\tanh \frac{E}{2k_B T}}{2E}, \tag{S28}$$

with the coupling constant $g$, another cutoff $E_0 \ll \varepsilon_0$, and the auxiliary density of states is given by

$$\nu_\gamma(E) = a_x a_y \int_{\varepsilon_0} \frac{dq_x dq_y}{(2\pi)^2} \gamma_{\boldsymbol{q}}^2\, \delta(E - \sqrt{\varepsilon_{\boldsymbol{q}}^2 + \Delta_{\boldsymbol{q}}^2}). \tag{S29}$$

It turns out that the auxiliary density of states $\nu_\gamma(E)$ for $\Delta = 0$ in the gap scenario (2) vanishes at the Van Hove point $r = 0$ as $\nu_\gamma(E) \propto E^2 \log 1/E$ for $E \to 0$. At zero energy, $E = 0$, and finite $r$ it assumes a finite value $\nu_\gamma(0) \propto r^2 \log 1/|r|$ for small $r$. As a result, the critical temperature $T_c(r)$ possesses a minimum at $r = 0$ and increases for increasing distance to the Van Hove point. As this behavior is inconsistent with the experimental signatures, we exclude this scenario (2) in the following.

For the numerical evaluation, we choose for the cutoffs $\varepsilon_0 = 0.1$ eV and $E_0 = 0.01$ eV. Moreover, we demand that the critical temperature $T_{c,\max} = 3.5$ K, i.e., $k_B T_{c,\max}/\varepsilon_0 \approx 0.003$ at the Van Hove singularity, $r = 0$, in order to connect with experiment. The coupling constant $g$ is chosen accordingly. The resulting phase diagrams for the scenarios (1) and (3) in the $(r, T)$ plane are shown in Fig. S12.

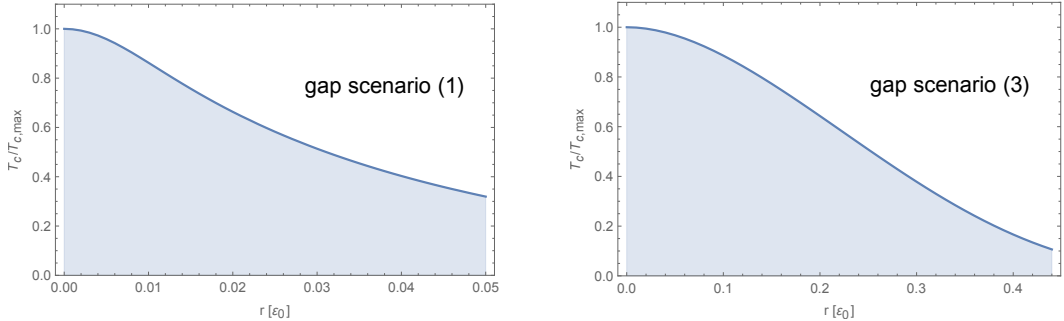

Figure S12. Phase diagrams in the $(r, T)$ plane for the gap scenarios (1) and (3) with a common $T_{c,\text{max}} = 3.5$ K. Note the different scales for the tuning parameter $r$ on the $x$-axis. The gap scenario (2) possesses a critical temperature $T_c(r)$ that is minimal at $r = 0$ and, as a consequence, it is not considered further.

3. *Entropy quench across the superconducting transition*

Here, we discuss the behavior of entropy close to the top of both superconducting domes in Fig. S12. The entropy per unit cell and per spin is given by

$$S(T, r) = -2k_B \int_0^\infty dE\, \nu(E) \Big( f(E) \log f(E) + (1 - f(E)) \log(1 - f(E)) \Big). \tag{S30}$$

It is symmetric $S(T, r) = S(T, -r)$ and we focus thus on positive $r \geq 0$. The entropy as a function of $r$ is evaluated for a few temperatures in Fig. S13.

The panels show the entropy divided by temperature because in the Fermi liquid limit this ratio is proportional to the density of states of the normal state, see also Eq. S12. This limiting behavior is expected for $T > T_c(r)$ and $r \gg k_B T$. As $k_B T_{c,\text{max}} \approx 0.003\, \varepsilon_0$ this amounts to $r/\varepsilon_0 \gg 0.003$ for $T \approx T_{c,\text{max}}$. Due to the different scales for the tuning parameter on the $x$-axis, this limit is mostly fulfilled in Fig. S13 for scenario (3) (provided that $T > T_c(r)$) but not yet for scenario (1).

As the superconducting dome is entered as a function of decreasing $r$, the entropy for the fully gapped scenario (1) exhibits a characteristic kink where the entropy changes slope $\partial_r S$ even close to the top of the dome for $T = 0.95\, T_{c,\text{max}}$. As a consequence, the maximum in $S(r)$ at $r = 0$ for $T > T_{c,\text{max}}$ is rapidly converted into a minimum for $T < T_{c,\text{max}}$. The behaviour for the nodal gap scenario (3) is distinctly different. As its density of states at the Van Hove point $r = 0$ is only capped by superconductivity but remains gapless, see Fig. S11, the entropy remains basically unquenched at $r = 0$ upon entering the superconducting dome. A small quench only appears for finite $r$ at $T < T_c(r)$ when the density of states develops a soft gap. This small quench is however not able to change the sign of the slope $\partial_r S$ at least not down to temperatures of $T = 0.6\, T_{c,\text{max}}$.

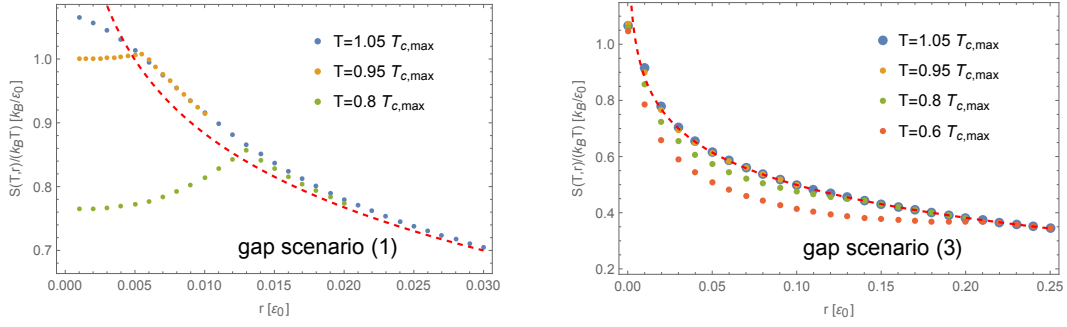

Figure S13. Entropy divided by temperature close to the top of the superconducting dome for scenarios (1) and (3) as a function of the tuning parameter $r$. The red dashed line corresponds to $S = \frac{\pi^2}{3} k_B^2 T \nu(0)|_{\Delta=0}$ and is proportional to the density of states in the normal state, see also the discussion in the context of Eq. S12.

Close to the Van Hove singularity, $r = 0$, the strain sensitivity is mostly attributed to the strain dependence of the tuning parameter, $r = r(\varepsilon_{xx})$. As a consequence, the slope $\partial_r S(r)$ is basically proportional to the strain derivative $\partial_{\varepsilon_{xx}} S$. A sign change of $\partial_r S$ is thus directly reflected in a sign change of the elastocaloric effect. The behavior for the fully gapped scenario (1) is, as expected, consistent with the calculation of section II B. It is also consistent with the experimentally observed signatures whereas the nodal gap scenario is not. So we conclude that the conversion of the maximum in entropy $S(\varepsilon_{xx})$ into a minimum upon entering the superconducting dome at its top as derived from measurements of the elastocaloric effect on $Sr_2RuO_4$ strongly suggests a superconducting order with a full gap at the Van Hove point in momentum space.